\documentclass[fleqn,usenatbib]{mnras}
\usepackage{newtxtext,newtxmath}
\usepackage{pdflscape} 
\usepackage{afterpage}

\usepackage[T1]{fontenc}
\usepackage{ae,aecompl}

\usepackage{graphicx}	
\usepackage{amsmath}	
\usepackage{amssymb}	
\usepackage[usenames,dvipsnames]{xcolor}
\usepackage[export]{adjustbox} 

\defcitealias{FFtech}{Paper~I}
\defcitealias{Hopwood15}{Hop15} 
\defcitealias{FFredshift}{Paper~II}
\defcitealias{FFncc}{Paper~IV}

\title[SPIRE FF III: Line ID and Off-Axis]{
The \textit{Herschel} SPIRE Fourier Transform Spectrometer Spectral Feature Finder III.
Line Identification and Off-Axis Spectra\thanks{\textit{Herschel} was an ESA space observatory with science instruments provided by European-led Principal Investigator consortia and with important participation from NASA.}}

\author[C. S. Benson et al.]{
Chris S. Benson,$^{1}$\thanks{E-mail: chris.benson@uleth.ca}
N. H\l{}adczuk,$^{2,3}$
L.~D. Spencer,$^{1}$
A. Robb,$^{1}$
J. Scott,$^{1}$
\newauthor
I. Valtchanov,$^{4}$
R. Hopwood,$^{4,5}$
and D.~A. Naylor.$^{1}$
\\
$^{1}$Institute for Space Imaging Science, Department~of Physics \& Astronomy, University of Lethbridge, 4401 University Drive, \\
Lethbridge, Alberta, T1K 3M4, Canada \\
$^{2}$ European Space Astronomy Centre, ESA, Camino Bajo del Castillo, 28692, Villanueva de la Ca$\tilde{n}$ada, Madrid, Spain \\
$^{3}$ Gran TeCan, S.A., Instituto de Astrof\'{i}sica de Canarias, C/ V\'{i}a L\'{a}ctea, S/N, 38205 - San Crist\'{o}bal de La Laguna, S/C de Tenerife, Spain\\
$^{4}$Telespazio Vega UK for ESA, European Space Astronomy Centre, Operations Department, 28691 Villanueva de la Ca\~nada, Spain\\
$^{5}$ Department of Physics, Imperial College London, Prince Consort Road, London SW7 2AZ, UK\\
}

\date{Accepted Jun.\ 2020. Received Jun.\ 2020; in original form Feb.\ 2020}

\pubyear{2020}

\begin{document}
\label{firstpage}
\pagerange{\pageref{firstpage}--\pageref{lastpage}}
\maketitle

\begin{abstract}
The ESA \textit{Herschel} Spectral and Photometric Imaging Receiver (SPIRE) Fourier Transform Spectrometer (FTS) Spectral Feature Finder (FF) project is an automated spectral feature fitting routine developed within the SPIRE instrument team to extract all prominent spectral features from all publicly available SPIRE FTS observations. We present the extension of the FF to include the off-axis detectors of the FTS in sparsely sampled single-pointing observations, the results of which have been ingested into the catalogue. We also present the results from an automated routine for identifications of the atomic/molecular transitions that correspond to the spectral features extracted by the FF. We use a template of 307 atomic fine structure and molecular lines that are commonly found in SPIRE FTS spectra for the cross-match. The routine makes use of information provided by the line identification to search for low signal-to-noise ratio features that have been excluded or missed by the iterative FF. In total, the atomic/molecular transitions of 178\,942 lines are identified (corresponding to 83\% of the entire FF catalogue), and an additional 33\,840 spectral lines associated with missing features from SPIRE FTS observations are added to the FF catalogue.
\end{abstract}




\section{Introduction}

The \textit{Hershcel Space Observatory$^\ast$} is an observatory class mission of the European Space Agency (ESA) \citep{Pilbratt10} that completed four years of observations exploring the far-infrared (FIR) and submillimeter (sub-mm) Universe in April 2013 with the depletion of its liquid cryogens \citep{herschelEnd}. The Spectral and Photometric Imaging REceiver (SPIRE) was one of three focal plane instruments on board \textit{Herschel}, consisting of both an imaging photometric camera and an imaging Fourier Transform Spectrometer (FTS) \citep{Griffin10}. The SPIRE FTS has two detector arrays, the Spectrometer Long Wavelength (SLW) and the Spectrometer Short Wavelength (SSW), that simultaneously cover a frequency band of 447--1546 GHz (SLW: 447--990 GHz, SSW: 958--1546 GHz). SPIRE FTS observations provide a wealth of molecular and atomic fine structure spectral lines that occur at FIR frequencies and provide measurements of the physical processes that occur in the interstellar medium (ISM) that give rise to star formation, within our own Galaxy and in more distant galaxies \citep{herschelGalaxies}. During \textit{Herschel's} mission, the SPIRE instrument provided 1\,349 high resolution ($\Delta \nu \sim 1.2$ GHz) spectral observations that are publicly available through the \textit{Herschel Science Archive} (HSA)\footnote{\url{http://archives.esac.esa.int/hsa/whsa/}} (see \citealt{FFtech}; \citealt{spire_handbook}).

Recently the SPIRE FTS observations have become more accessible (particularly to those unfamiliar with the instrument) through the SPIRE Spectral Feature Finder Catalogue\footnote{\url{https://www.cosmos.esa.int/web/herschel/spire-spectral-feature-catalogue}}, which includes a collection of significant spectral features extracted from all publicly available high resolution (HR) single-pointing and mapping observations by the automated SPIRE Feature Finder (FF) routine \citep{FFtech}\footnote{Due to repeated referencing of \citet{FFtech} and \citet{FFredshift}, the abbreviations \citetalias{FFtech} and \citetalias{FFredshift} are used in further text of this paper.}. The full SPIRE Automated Feature Extraction Catalogue (SAFECAT), in its second full release, contains the central frequency and signal-to-noise ratio (SNR) of 167\,525 features at SNRs greater than five from all publicly available SPIRE FTS observations from the HSA. From the spectral content of each observation and through literature cross references, radial velocity estimates for each observation are also catalogued \citep{FFredshift}.

The FF, as presented by \citetalias{FFtech}, only considers the central detectors of the SLW and SSW detector arrays (SLWC3 and SSWD4) for sparsely sampled single-pointing observations (hereafter referred to as \textit{sparse} observations). Most observations of this type are of sources that are point-like or have very little spatial extent in comparison to the SPIRE FTS beam (\citetalias{FFtech}, \citealt{Makiwa2013}, \citealt{Wu2013}). During the development of the FF, we have identified 357 observations of sources which have large spatial extent and are expected to have significant spectral information measured by the other 52 detectors that are off of the optical axis \citepalias{FFtech}. In a sparse observation of the crab nebula, \citet{argon} discovered the first detection of the ionic molecule $^{36}$ArH$^+$ in an astronomical source in the spectrum from an off-axis detector. 

A template of spectral features that are expected within sources related to SPIRE science goals has been identified. This template is used to match the results from the FF routine to specific atomic fine-structure and molecular spectral features and, in conjunction with SAFECAT, provides initial spectral assignment of the atomic and molecular composition in SPIRE FTS observations. This is to serve as a data mining aid for exploitation of the HSA. 

The scope of this paper is to report on the application of the FF to the spectra of off-axis detectors in \textit{sparse} SPIRE FTS observations and present the identification of the molecular/atomic transitions corresponding to each of the spectral features contained in SAFECAT. These are included in the FF catalogue. Also included is a description of a routine to work in conjunction with the FF to improve the completeness of the catalogue, particularly for low SNR lines. Our goal through this work is to further improve on the completeness of SAFECAT and, through line identification, provide astronomers with a useful tool for expedited access and scientific exploitation of the entire SPIRE FTS catalogue of observations. This paper is one of a series of four which discuss aspects of the FF: the main technical paper \citep[\citetalias{FFtech}:][]{FFtech}, a radial source velocity paper \citep[\citetalias{FFredshift}:][]{FFredshift}, a line identification paper (Paper III: this paper) which also presents FF results for the off-axis spectra within sparse observations, and a [CI] detection and deblending paper \citep[\citetalias{FFncc}:][]{FFncc}.

\section{Application of the FF to Off-Axis Detectors}
\label{sec:offAx}
\begin{figure}
	\includegraphics[trim = 8mm 5mm 7mm 2mm, clip, width=1.0\columnwidth]{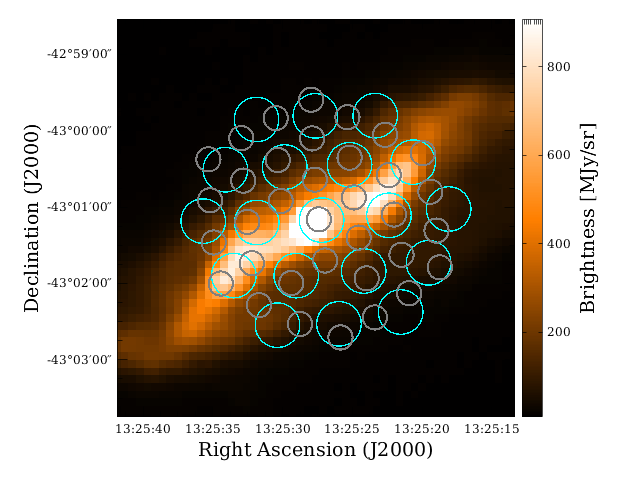}
	\hspace{-1.2pt}\includegraphics[trim = 8mm 5mm 7mm 2mm, clip, width=1.0\columnwidth]{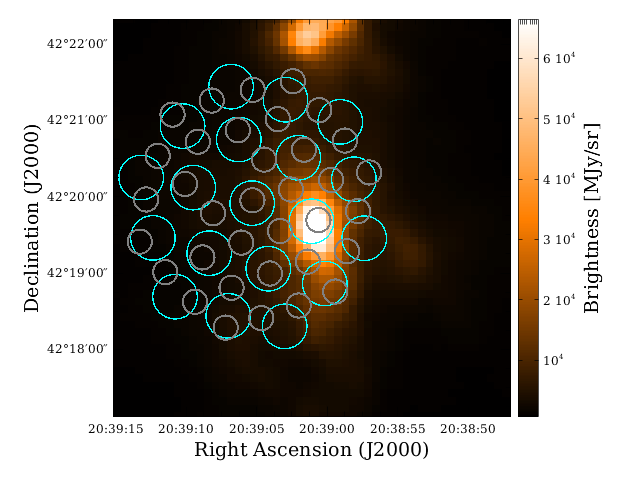}
    \vspace{-12pt}\caption{Two examples of significant emission in off-axis detectors. The footprints of the SLW array (cyan) and SSW array (grey) are superimposed over photometer maps taken by the SPIRE photometer short wavelength (PSW) array (250 $\mu$m). The top panel is an observation of the Seyfert Galaxy Centaurus A (observation ID 1342204037) showing extended emission across the SPIRE FTS footprint and the bottom panel is an observation of L1448 (observation ID 1342202264) showing prominent emission in an off-axis pixel.}
    \label{fig:photFoots}
\end{figure}
The FF as described by \citetalias{FFtech} only considers spectra from the central detectors of sparsely sampled single-pointing observations but there are several cases in SPIRE FTS observations where these observations will have important spectral information in off-axis detectors. \citetalias{FFtech} have identified 357 observations of semi-extended or fully extended sources that are expected to have significant emission in the territory of off-axis detectors (See Fig.\,\ref{fig:photFoots}). These extended or semi-extended observations account for 41.1\% of unique sparse observations. There is also a known potential for pointing errors in SPIRE FTS observations that can sometimes result in an observation straying as much as 1 arcminute from the requested pointing \citep{SanchezPortal2014} resulting in observations where the brightest emission is not seen by the central detector. An extreme case of this pointing offset is shown in the bottom panel of Fig.\,\ref{fig:photFoots}. The photometer map is centered on the requested pointing while the footprint of the SPIRE FTS is shown in its observational position, $\sim1$ arcminute from the requested pointing. 

\begin{figure}
	\includegraphics[trim = 0mm 0mm 0mm 0mm, clip, width=\columnwidth]{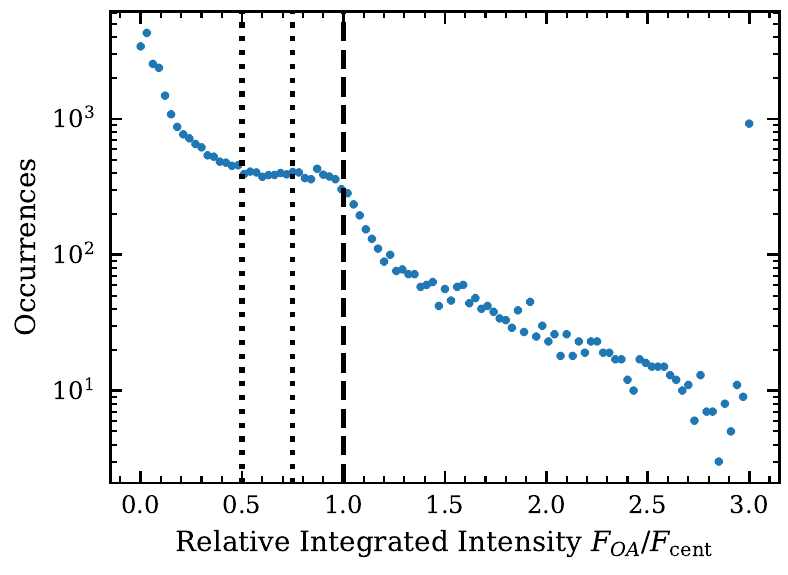}
    \vspace{-17pt}\caption{The integrated intensity of the spectra from off-axis detectors, $F_{OA}$, compared to the integrated intensity of the central detector in the same observation $F_{\text{cent}}$. Vertical black lines mark the regions of spectra with more than 50\%, 75\% and 100\% of the integrated intensity of the central detector (from left to right), which represent 32.4\%, 22.6\%, and 13.1\% of the data, respectively.}
    \label{fig:offAxIntensHist}
\end{figure}

The number of SPIRE FTS observations that are expected to have significant emission in off-axis pixels can be estimated by comparing the integrated intensity across the entire SPIRE band. We have found that 26.4\% of sparse observations have at least one off-axis SLW detector that is brighter than that of the central detector. For the SSW array, 25.1\% of observations have a brighter off-axis detector than the central detector. The integrated intensity of each individual off-axis detector from all 868 unique sparse observations of the SPIRE FTS, $F_{OA}$, compared to the integrated intensity from the central detector, $F_{\text{cent}}$, of their respective observation is shown as a histogram in Fig.\,\ref{fig:offAxIntensHist}. 

\subsection{Feature Finder Results}
\begin{figure}
	\includegraphics[trim = 0mm 3mm 0mm 2mm, clip, width=\columnwidth]{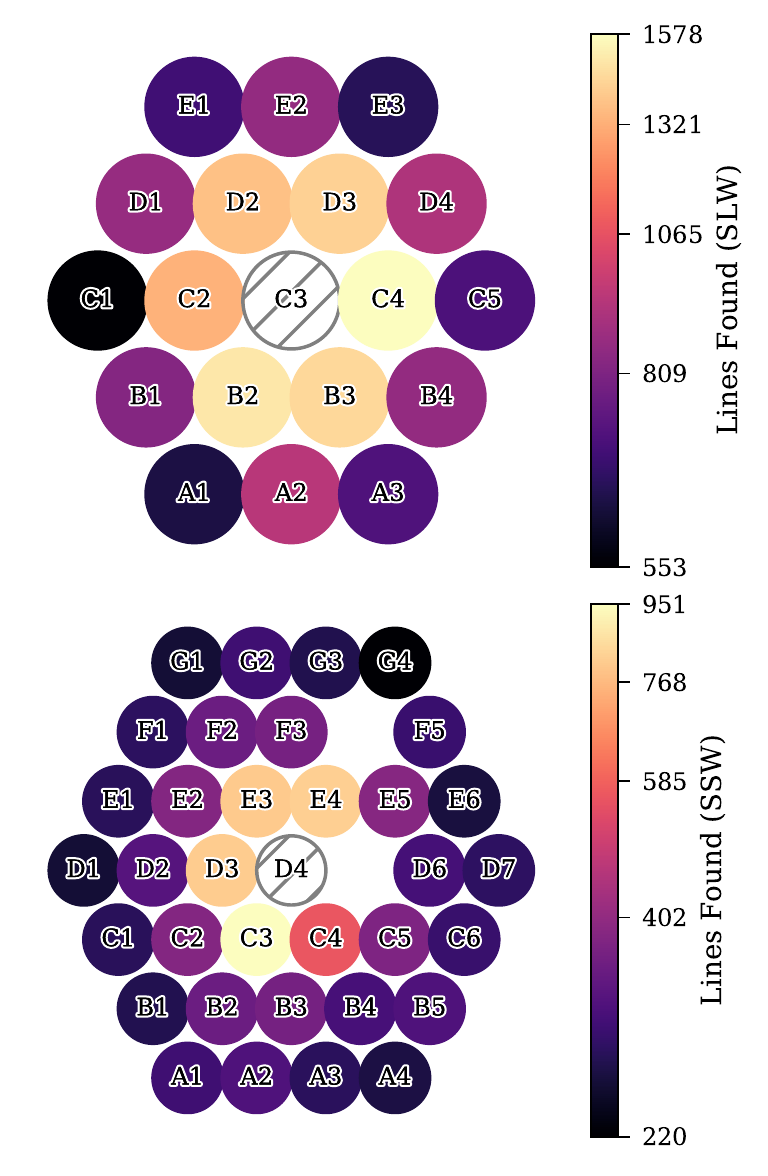}
	\vspace{-12pt}\caption{The number of lines detected in each off-axis detector by the FF. The SLW array is shown on top while the SSW detector array is shown underneath.} 
	\label{fig:DetDetections}
\end{figure}
Applying the standard FF routine to the off-axis detectors of all 868 sparse observations yielded a total of 30\,720 additional lines with an $|$SNR$|$ greater than 6.5. The results from the FF are broken down per detector in Fig.\,\ref{fig:DetDetections}. As expected, the majority of lines discovered are close to the central detector of each array. An $|$SNR$|$ cutoff of 6.5 for emission features and 10 for absorption features is used to avoid spurious detections. Nominally a cutoff of $|$SNR$|$ 5 is employed for emission features in spectra from the central detectors while a cutoff of 10 is used for absorption features \citepalias{FFtech}. The removal of possible spurious detections from off-axis spectra discards 36.3\% of the features found. 

An additional observation of the calibration source, AFGL\,4106 (\citealt{Hopwood15}\footnote{Due to repeated referencing of \citet{Hopwood15}, the abbreviation \citetalias{Hopwood15} is used in further text of this paper.}), that is nominally excluded from the FF catalogue for central detectors is considered by the FF for off-axis detectors \citepalias{FFtech}. This observation (ID 1342208380) suffers from an exceptionally large pointing error of $\sim$\,24 arcseconds causing the source to be missed by the central SSWD4 detector and to be only partially observed by the SLWC3 detector. Treatment of the spectrum from the SLWC3 detector in this observation requires an extensive pointing offset correction which was deemed beyond the scope of the FF. The central source is not observed by any of the off-axis detectors but these detectors do provide measurements of the surrounding emission from the galactic cirrus that do not require any such corrections.

\begin{figure}
	\includegraphics[trim = 0mm 0mm 0mm 0mm, clip, width=\columnwidth]{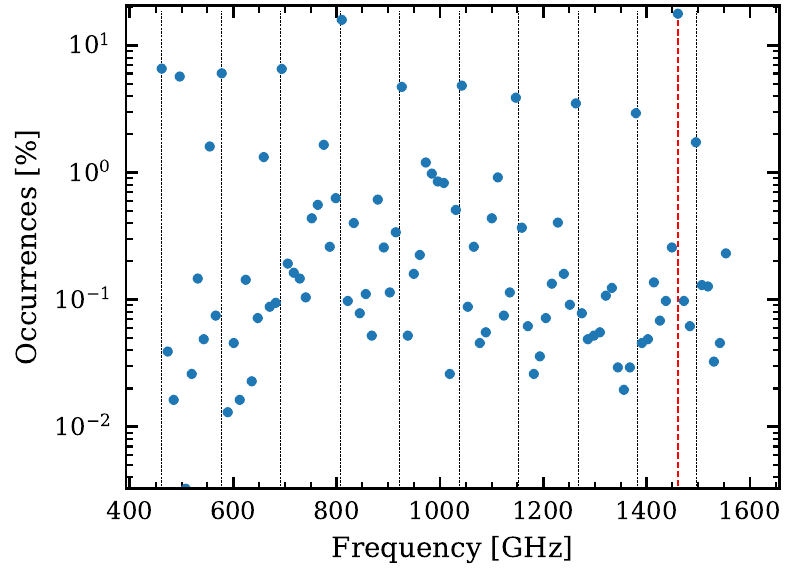}
	\vspace{-17pt}\caption{The number of lines detected by off-axis detectors at different frequencies. Histogram bin widths are 11.6\,GHz wide. The rest frequencies of $^{12}$CO are marked by vertical black lines while the [\ion{N}{II}] $^3$P$_1$--$^3$P$_0$ transition is marked by the dotted red line.} 
	\label{fig:freqHist}
\end{figure}

The number of features found at different frequencies in the SPIRE FTS band is shown in Fig.\,\ref{fig:freqHist}. We see that the [\ion{N}{II}] $^3$P$_1$--$^3$P$_0$ feature (marked by the vertical red line) is frequently detected in off-axis detectors accounting for over 10\% of the total detected features. $^{12}$CO features (marked in grey) are also fairly common, as expected (see \S\,\ref{sec:lineID}). 
We found that of the spectra from off-axis detectors with features found by the FF, 3\,590 of 27\,673 (13.0\%) have reliable velocity estimates based on the $^{12}$CO features (see \citealt{FFredshift}\footnote{For \citet{FFredshift}, the abbreviation \citetalias{FFredshift} is used in further text of this paper.}). This is significantly less than the success rate for the central detectors which suggests that a significant number of off-axis spectra are dimmer and contain more spurious detections. In order to improve line identification in off-axis detectors we extended the cross-correlation method (see \citetalias{FFredshift}) to measure the radial velocity of the spectra from off-axis detectors. This increases the number of spectral observations with reliable velocity estimates from 3\,590 to 9\,211 (33.3\%).

There exist 3\,110 cases where a velocity estimate is provided by both the $^{12}$CO and cross-correlation routines allowing us to further the validation of both routines performed by \citetalias{FFredshift} by including the results from off-axis detectors of the FTS. We found that 92.4\% of velocity estimates  provided by both routines are in agreement to within 20\,km/s, which is similar to the agreement for central detectors \citepalias{FFredshift}. Additionally, we have found that the neutral carbon routine designed within the FF to improve its ability to disentangle the closely spaced ($\sim$\,2.7\,GHz) $^{12}$CO(7-6) and [\ion{C}{I}] $^3$P$_2$--$^3$P$_1$ features has successfully contributed 215 spectra features to the total features from off-axis detectors corresponding to $^{12}$CO(7-6), [\ion{C}{I}] $^3$P$_2$--$^3$P$_1$, and/or, [\ion{C}{I}] $^3$P$_1$--$^3$P$_0$ transitions (see \citealt{FFncc}).

There are several sparsely sampled single-pointing observations that have a bright central FIR source with strong molecular emission and $[$\ion{N}{II}$]$ across the entire field viewed by the FTS. Fig.\,\ref{fig:NIIdecouple} shows an example of this $[$\ion{N}{II}$]$ emission surrounding the calibration source AFGL 4106 (see \citetalias{Hopwood15} and \S\,\ref{sec:calSources} of this paper). In this case the $[$\ion{N}{II}$]$ emission is useful for decoupling the dusty star measured by the central detector from the surrounding Galactic cirrus which has a radial velocity that differs by as much as $\sim 60$\,km/s.

\vspace{-6pt}
\subsection{Postcards}
\label{sec:postcards}
The postcards developed to highlight off-axis spectral content extend the previously developed sparse postcards for central detectors and have been designed to follow a similar format to the mapping postcards described by \citetalias{FFtech}. The postcards are composed of 8 panels, an example of a postcard is shown in Fig.\,\ref{fig:offAPost}. The goal of FF postcards is to provide an at-a-glance summary of the information provided by the FF for any given observation. It should be noted that the object names reported in SPIRE observations are submitted by the observer and do not necessarily conform to any naming standards. For the convenience of generating postcards in mass quantities, the reported names in postcards are those that are directly recorded in the metadata of HSA observations as provided by the original observation investigators.

The first of the four columns of the postcard displays the sparse postcard that was generated in the nominal FF process (top) and the SPIRE FTS footprint over the PSW photometer map (bottom). The sparse postcard shows the spectra from each central detector (SLWC3 in red, SSWD4 in blue) and vertical shaded regions mark the 10\,GHz wide edges of each band, which are nominally avoided by the FF in line searching. Features extracted by the FF are marked by vertical sticks whose length is proportional to the SNR of the line. The fitted continuum polynomial is also shown in green. For the photometer map, the figure itself is centred on the requested pointing of the observation and cropped to a 4 arcminute square. The colourbar of the photometer map is scaled to reflect the image content and empty pixels which are assigned a not-a-number (NaN) value are coloured in grey. The SPIRE footprint is shown with its actual pointing for each detector, the SLW array is outlined in cyan and the SSW array is outlined in grey. The circles represent the FWHM of the detector beam at the centre of the band corresponding to each array, $\sim$25 arcseconds for SLW and $\sim$19 arcseconds for SSW \citep{Makiwa2013, spire_handbook}. In cases where no SPIRE photometer map is available for the pointing of the FTS observation, a stock image stating that no photometer map was found is shown instead. There are 97 SPIRE FTS observations for which no photometer map was observed.

\afterpage{
\begin{landscape}
\begin{figure}
	\vspace{48pt}
	\includegraphics[width=\columnwidth]{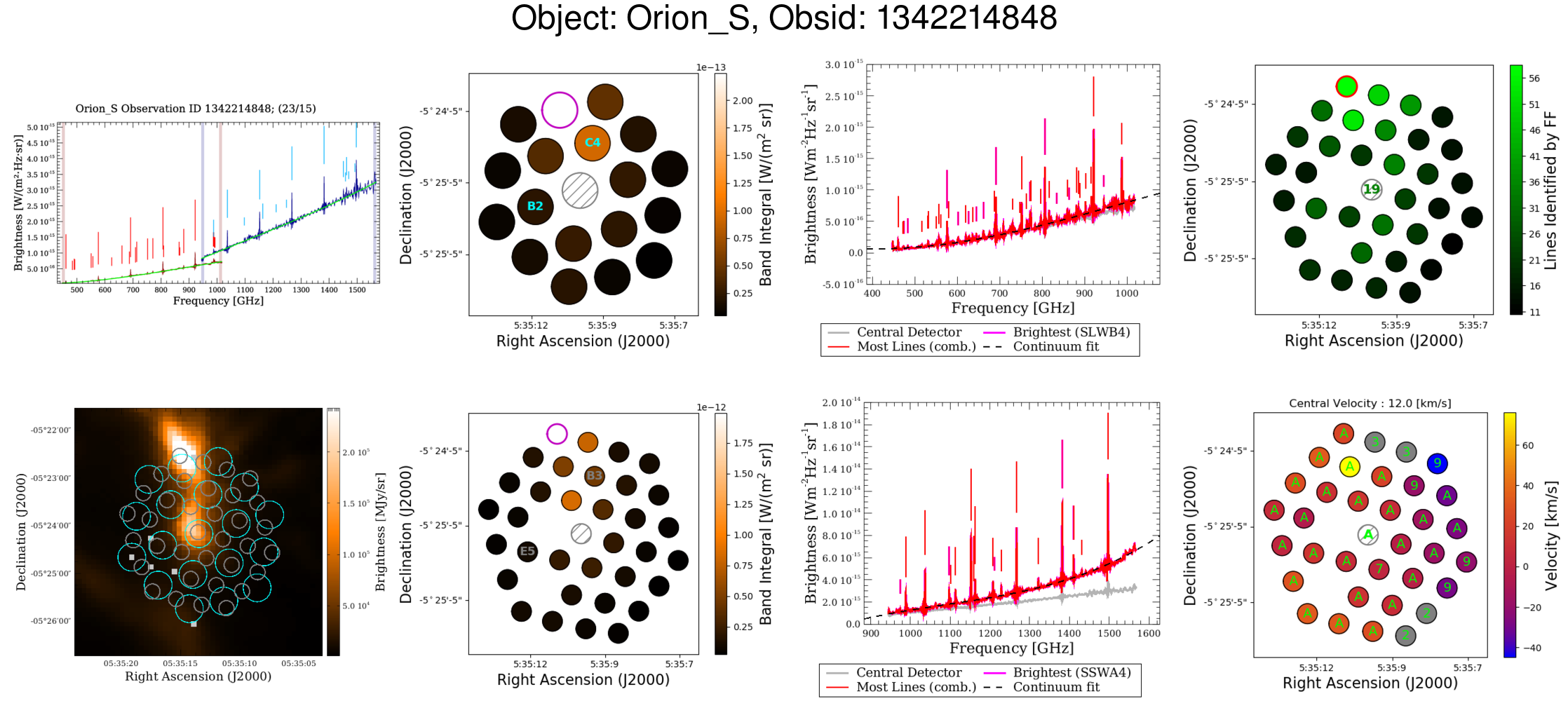}
    \caption{A sample off-axis postcard (see \S\,\ref{sec:postcards}). The first column shows the FF postcard from the central detectors (top) and the footprint of the FTS observation (SLW cyan, SSW grey) imposed on the SPIRE PSW photometer map (bottom). The second column presents the integrated flux across the entire SLW (top) and SSW (bottom) bands. The third column demonstrates the spectra from the brightest off-axis detector, the detector corresponding to the most features found by the FF (outlined in the same colour in the adjacent plots), and the central detector, plotted in magenta and grey, respectively, for SLW (top) and SSW (bottom). The fourth column shows the number of lines found by the FF (top) and the radial velocity (bottom) estimates from the $^{12}$CO routine. In this column only SSW detectors are shown but the number of lines and velocity estimates are determined by combining each SSW detector with its closest SLW detector.}
    \label{fig:offAPost}
\end{figure}
\clearpage
\end{landscape}
}
\afterpage{
\begin{landscape}
\begin{figure}
	\centerline{
    \vspace{48pt}
    \includegraphics[width=\columnwidth]{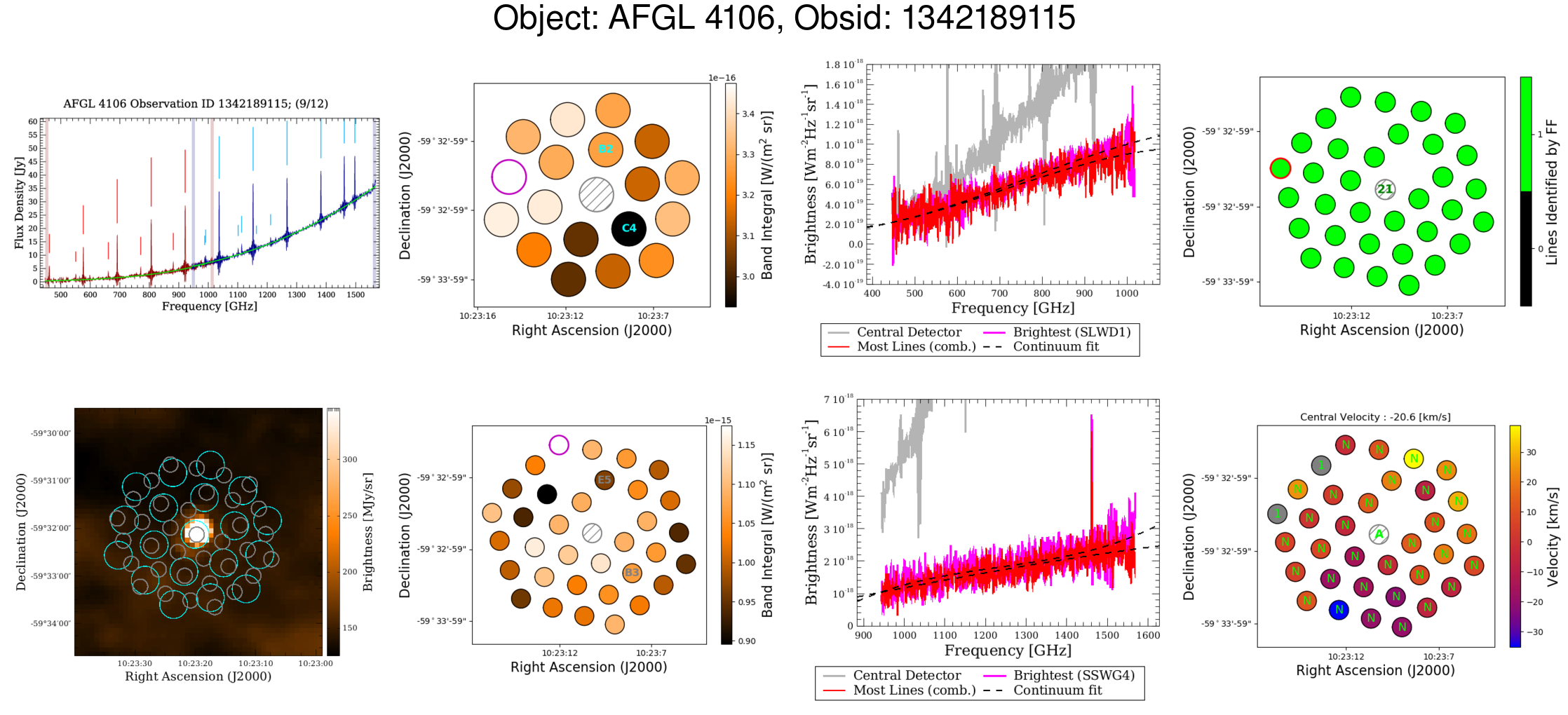}}
	\caption{The off-axis postcard for an observation of the post-red supergiant AFGL 4106. The dusty molecular cloud surrounding the star is observed by the central detector while ionized nitrogen from the surrounding Galactic cirrus can be seen in the off-axis detectors.} 
	\label{fig:NIIdecouple}
\end{figure}
\end{landscape}
}

The second column contains the footprint of each detector array on sky (SLW top, SSW bottom). Each off-axis detector is coloured based on the integrated intensity across their entire band for the given observation, with the extended source calibrated spectrum used for this calculation. The central detector is not included in the determination of the colourbar dynamic ranges and it is marked by the grey hashed-out circle to illustrate this. In order to determine the orientation of the detector arrays, the SLWC3 and SLWB2 detectors are labelled in cyan while the overlapping SSW detectors, SSWB3 and SSWE5, are marked in grey. Note the two dead detectors in the SSW array which also help in determining the orientation of the FTS on-sky. The brightest detector from each array is outlined in magenta. This intensity map of the FTS footprint with the photometer map provides information about the structure of the source which should correspond to the spectral content of off-axis detectors. 

The third column presents a few spectra of interest from the SLW (top) and the SSW arrays (bottom). The spectra from the brightest off-axis detectors are plotted in magenta (based on the band integrated intensity and marked in the corresponding panels to the left) and the spectra from the detectors with the most lines are plotted in red (marked in the upper panel to the right). The spectrum from the central pixel is also plotted in the figure background, in grey, and intentionally without influencing the vertical axis limits. The central spectrum is only provided for comparison and is better viewed in the central detector postcard (top left panel). Again the extended source calibration spectra are chosen for these plots. The spectra with the largest number of lines is determined by combining the lines extracted from each SSW spectrum with those from its nearest neighbouring SLW spectrum, the number of features is then determined as the total number of features from both. As is the case for central detector postcards, the lines extracted by the FF are marked by vertical sticks with lengths proportional to the SNR of the feature. Emission features are marked above the spectra while absorption features are marked below. 

The fourth and final column contains the number of lines extracted by the FF (top) and the radial velocity estimates (bottom) from off-axis detectors. In each case, the lines extracted from the SSW detectors are paired to the nearest SLW detector and used for velocity estimates and the total number of lines displayed in each SSW detector. The SSW off-axis detector with the most number of lines found by the FF is outlined in red. The number of spectral features found by the FF for the central detector is written in green on the central detector. The velocity estimates are obtained from the $^{12}$CO velocity estimate routine and follow similar annotations to those in mapping postcards. The green numbers mark the number of $^{12}$CO features used in the velocity estimate with the character `A' denoting that all 10 in-band $^{12}$CO features are used. When the [\ion{N}{II}] feature is used for the velocity estimate an `N' character is used instead. Detectors for which no reliable velocity estimate can be obtained by the $^{12}$CO routine are coloured in grey. The radial velocity estimate from the central pixel is also shown in the title of the velocity panel and the radial velocity flag is printed in green over the central pixel.

The FF results from the off-axis detectors and postcards for sparse SPIRE FTS observations are available through SAFECAT and as individual products for each observation \citepalias{FFtech}. These are stored in the ESA Herschel legacy area\footnote{\label{ff_legacy}The FF \textit{Herschel} legacy area is available at the following URL: \url{http://archives.esac.esa.int/hsa/legacy/HPDP/SPIRE/SPIRE-S/spectral_feature_catalogue/}}. This addition to the catalogue inherits the HSA searching tools and accessibility of the nominal catalogue detailed in \citetalias{FFtech}. 

\begin{figure}
	\includegraphics[width=\columnwidth]{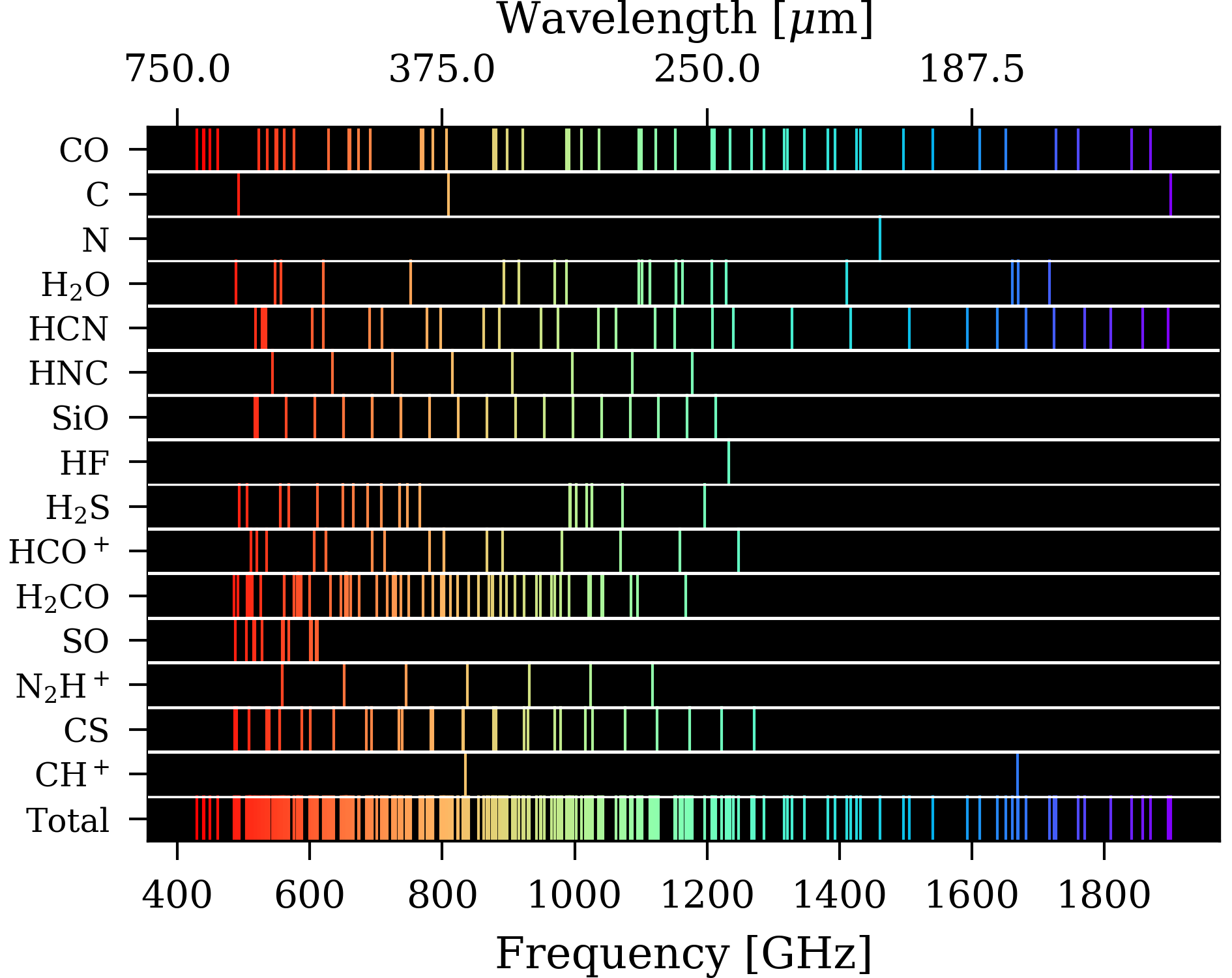}
    \vspace{-12pt}\caption{The collection of atomic fine-structure and molecular lines contained in the identification template at rest-frame. The false-colour is chosen such that the spectral feature in the template with longest wavelength is the most red.}
    \label{fig:template}
\end{figure}

\section{Line Identification}
\label{sec:lineID}
As one of the final steps in the \textit{Herschel} SPIRE feature finder routine, we attempt to determine the atomic/molecular transitions that correspond to the prominent FF spectral features. Through this work, SAFECAT will not only provide to users the central frequency and SNR of significant spectral features but also provide information concerning molecular/atomic composition of sources.

In order to identify the atomic/molecular transitions that correspond to the spectral features extracted by the FF, these features are compared to a template line list of 307 atomic fine-structure and molecular features that are commonly found in astronomical sources at far-infrared wavelengths (Fig.\,\ref{fig:template}). The template is predominantly composed of spectral features from the \textsc{CASSIS} software \citep{templateCassis}\footnote{\url{http://cassis.irap.omp.eu/?page=presentation}}, a free spectral analysis software designed to work with \textit{Herschel} that can be accessed through the Herschel Interactive Processing environment (\textsc{HIPE}) \citep{HIPE}. A number of features that are also commonly found SPIRE FTS spectra, such as the $[$\ion{N}{II}$]$ $^3$P$_1$--$^3$P$_0$ fine structure line, have been included in the template as well. The template makes use of the publicly available information contained in the following spectral databases: the NIST/Lovas Atomic/molecular Spectra Database \citep{NISTcat, NISTlovas}, the CDMS database \citep{CDMScat}, and the JPL sub-millimetre, millimetre and microwave spectral line catalogue \citep{JPLcat}. The full list of the lines used in the template is shown in Tab.\,\ref{tab:catalogue}. These spectral features correspond to 15 different atomic/molecular species and encompass transitions from different isotopologues, ionization states, and vibrational bands. These atomic/molecular species and the number of lines in the template corresponding to each are summarized in Tab.\,\ref{tab:TemplateSumm} and illustrated in false colour in Fig.\,\ref{fig:template}.

%
\begin{table}
\begingroup
\begin{center}
\newdimen\tblskip \tblskip=5pt
\caption{\label{tab:TemplateSumm} A summary of the template used in line identification. The full list including rest frequencies of each transition can be found in Tab.\,\ref{tab:catalogue}.}
\nointerlineskip
\small
%
\newdimen\digitwidth
\setbox0=\hbox{\rm 0}
\digitwidth=\wd0
\catcode`*=\active
\def*{\kern\digitwidth}
\newdimen\signwidth
\setbox0=\hbox{+}
\signwidth=\wd 0
\catcode`!=\active
\def!{\kern\signwidth}
%
\tabskip=2em plus 2em minus 2em
\halign to \hsize{\hfil#&\hfil*#\hfil& \hfil#\hfil& \hfil#\hfil& \hfil#*\hfil&#\hfil\cr
 &\multispan4\hrulefill& \cr
\noalign{\vspace{-8.0pt}}
 &\multispan4\hrulefill& \cr
& Molecule/Atom& Lines& Molecule/Atom& Lines& \cr
\noalign{\vspace{-5.5pt}}
 &\multispan4\hrulefill& \cr
&	$[$\ion{C}{I}$]$& *2&  $^{13}$C$^{17}$O& *4& \cr
&	$[$\ion{C}{II}$]$& *1& $^{13}$C$^{18}$O& *2& \cr
&	$[$\ion{N}{II}$]$& *1& N$_2$H$^+$&       *7& \cr
&	CH$^+$&            *2& HCO$^+$&          *9& \cr
&	p-H$_2$O&          *9& H$^{13}$CO$^+$&   *5& \cr
&	o-H$_2$O&          10& HC$^{18}$O$^+$&   *1& \cr
&	HDO&               *1& p-H$_2$CO&        35& \cr
&	H$_2^{18}$O&       *2& o-H$_2$CO&        40& \cr
&	HCN&               23& p-H$_2$S&         *7& \cr
&	H$^{13}$CN&        14& o-H$_2$S&         13& \cr
&	HNC&               *8& CS&               19& \cr
&	CO&                13& $^{13}$CS&        12& \cr
&	$^{13}$CO&         14& SiO&              19& \cr
&	C$^{17}$O&         *9& SO&               15& \cr
&	C$^{18}$O&         10& HF&               *1& \cr
\noalign{\vspace{-7.5pt}}
 &\multispan4\hrulefill& \cr
\noalign{\vspace{-7.5pt}}
}
\end{center}
\endgroup
\vspace{-12pt}
\end{table}

These molecular and atomic fine structure lines included in the template provide effective tools for measuring thermodynamic processes, the kinetic motion of shocks, and ionization processes in the ISM, and are of particular importance in FIR astrophysics. The CO molecule is the most abundant molecule in the Universe following H$_2$ \citep{tennysonAstro} and is the most important coolant in cold dense molecular clouds \citep{dysonISM, COcooling, molecCooling}. Due to its dipole moment, CO has several rotational transitions that radiate at FIR wavelengths. The CO molecule is easily excited into rotational energy states as it collides with H$_2$ in the cool ISM and it relaxes to ground state quickly \citep{dysonISM}. The lowest transition in the $^{12}$CO ladder ($J = 1$--$0$) corresponds to an energy transition equivalent to 5.5\,K meaning the molecule is an effective coolant down to this temperature \citep{dysonISM, COcooling}, thus CO provides an important tool for probing star-forming regions \citep{gibionThesis, tennysonAstro}. The SPIRE FTS was designed to cover rotational transitions of $^{12}$CO from $J=4$--$3$ to $J=13$--$12$. Water (H$_2$O) is an asymmetric rotor and several lines from gas-phase water vapour occur in the SPIRE FTS band. Like CO, these rotational transitions provide an important coolant in the ISM \citep{gibionThesis}. SPIRE measurements of water vapour are particularly useful for studying ultraluminous infrared galaxies \citep{waterULIRG1} and can be used as a tracer of star formation in the ISM \citep{waterISO}. Fine structure lines from $[$\ion{C}{I}$]$ and $[$\ion{N}{II}$]$ are important coolants for the gas phase of the ISM \citep{gibionThesis}. They are of particular import as they probe different energy regimes than the molecules included in the template \citep{FFncc, COcooling}. The utility of lines from molecules, atoms, and ionized atoms are not limited to nearby regions of the ISM. The utility of [\ion{C}{I}], [\ion{N}{II}], and CO features measured by the SPIRE FTS to probe the properties of nearby and distant galaxies has been shown by \citet{specExample2} and \citet{specExample}, respectively.
The line identification routine is applied to all 1\,025 HR SPIRE FTS observations that have lines identified by the FF and have a reported radial velocity estimate \citepalias{FFredshift} (note that sparse observations may have two FF products, one for each calibration. See \citetalias{FFtech} for more details). The template is matched to the features extracted by the FF routine by first shifting the frequencies of the FF features to the rest frame, then all template lines, which contain the atomic/molecular species and quantum numbers of each transition, that are within 0.3\,GHz (1/4 the resolution of an HR observation) of a FF feature are matched to that feature. 

Under Local Thermodynamic Equilibrium (LTE) conditions, the intensities of spectral lines corresponding to a single atomic/molecular species follow a Boltzmann distribution \citep{dysonISM, hollas2004modern} and it is expected that when one spectral line for a given atomic or molecular species is found within the SPIRE band the other transitions corresponding to this species within the band, and LTE regime, will also be present. It is important to note that many astronomical systems are not in LTE \citep{dysonISM}. For the scope of this work, the expectation of finding multiple lines corresponding to a single species (provided they are within the band) is a useful assumption. 

When multiple template features are matched to a single FF extracted line, all possible matches are reported in the catalogue and each is compared against the other identified lines to determine which of the possible atomic/molecular species has the most other identified lines. The most likely line of the multiple matches based on this criteria is then marked as the most plausible identification. If multiple species have the same number of identified lines, both features are marked as plausible.


\subsection{Low SNR lines}
\label{sec:lowSNR}
\begin{figure}
	\includegraphics[trim = 5mm 4mm 1mm 1mm, clip, width=\columnwidth]{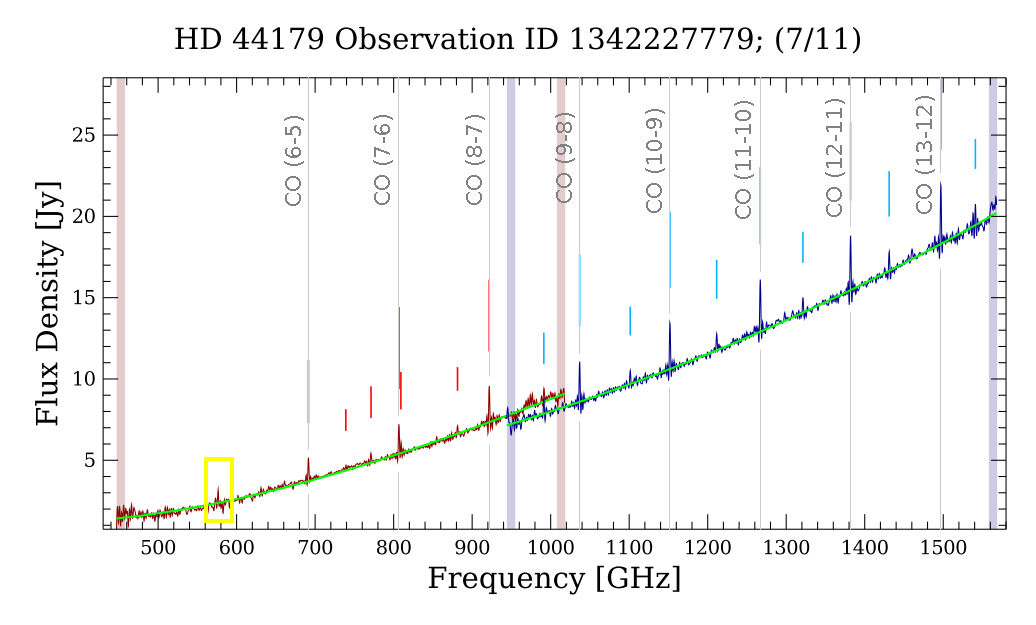}
    \vspace{-16pt}\caption{The SPIRE spectral postcard from SAFECAT of the post-AGB star HD 44179. The yellow box emphasizes the CO $J=5$--$4$ transition excluded from the initial pass of the FF that is found by the second pass looking for low SNR features using the other identified transitions in the CO ladder.}
    \label{fig:missingLine}
\end{figure}

The main FF catalogue does not report features with a $|$SNR$|$ less than five in order to avoid the inclusion of spurious detections \citepalias{FFtech}. It is easy to imagine a situation where an atomic or molecular species containing multiple lines in the SPIRE FTS band has some features detected near this limit with some actually falling below it. In the main execution of the FF these features would not be included in the catalogue. One such example is shown in Fig.\,\ref{fig:missingLine}. Strong $^{12}$CO emission is evident in this observation of the post-Asymptotic Giant Branch (AGB) star HD 44179, but the $J=5$--$4$ transition is not included in the FF catalogue since it occurs at a SNR less than 5 (yellow box of Fig.\,\ref{fig:missingLine}).

The line identification obtained by matching with the template provides a useful tool for extracting low SNR lines that are not nominally reported by the FF. A check for low SNR features is done by iterating through the atoms/molecules identified in an observation looking for transitions that are expected to be found in the spectrum but are not contained in the FF results. This list of missing lines is then inserted into a second iteration of the FF routine as an initial central frequency guess for additional features during the final (and lowest) SNR threshold. Using this method we search for additional features down to $|$SNR$|= 2$ to add to the catalogue. In the case of HD 44179 in Fig.\,\ref{fig:missingLine}, the missing CO $J=5$--$4$ is found by this low SNR search at 576.35\,GHz with an SNR of 4.35. Similarly the $^{13}$CO $J=5$--$4$ transition was found in this source at 661.00\,GHz with an SNR of 4.45.

\begin{figure}
	\centerline{\includegraphics[width=\columnwidth]{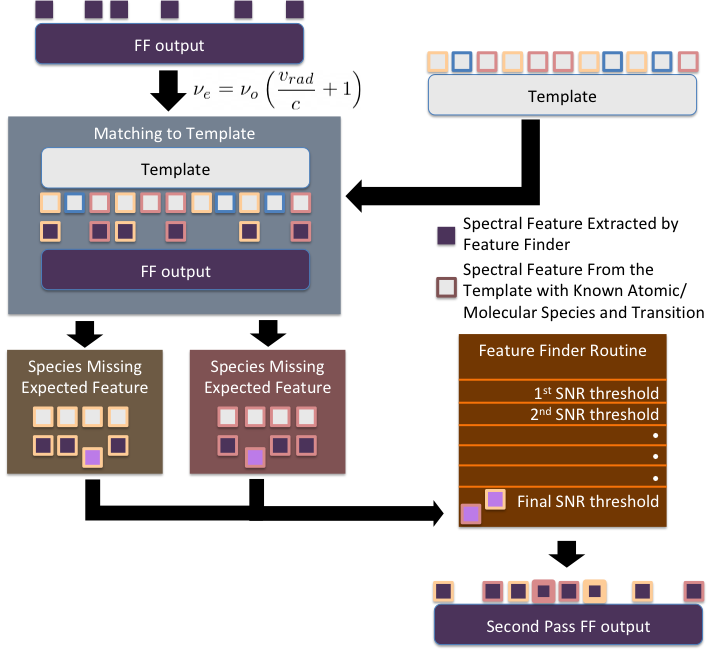}}
	\vspace{-4pt}\caption{A flow chart describing the line identification process and how it is used to find low SNR lines in the FTS spectra that may have been excluded from FF catalogue. A full step-wise description of the algorithm is presented in the text.}
	\label{fig:lowSNRFlow}
\end{figure}
The line identification process and low SNR search routine are summarized by the flow chart in Fig.\,\ref{fig:lowSNRFlow}:
\begin{enumerate}
\item Features output by the FF from a spectrum with a radial velocity estimate (shown as purple squares in the flow chart) are shifted into the rest frame velocity.
	\item Features are compared to the template (shown as grey squares) where  atoms/molecules, and transitions, of the FF features are identified.
	\item The low SNR search routine iterates through the identified species and determines if any transitions are missing (shown by the lighter coloured squares). These missing expected features are added to a list.
	\item The FF routine is repeated on the SPIRE FTS observation from the HSA. During the final iteration of SNR thresholds, which is changed to include features $5>|$SNR$| \geq 2$, frequencies shifted to the velocity frame from the list of missing features are used as initial guesses for the central line frequency of features to be fit by sinc functions. These features are assumed to be of low SNR.
	\item The new list of features output by the FF is subjected to the nominal checks for removing features in the FF \citepalias{FFtech} with the exception of the lower SNR threshold and the masks around prominent features that have previously been fit (for missing expected features that may lie close to another high SNR line).
	\item Remaining features within the list of missing but expected features are recorded and flagged (shown with the thicker outline).
	\item Features found by the low SNR search routine are appended to the output of the line identification routine with a `!' character in the comment section.
\end{enumerate} 

The low SNR search routine has also been found to discover features that occur above the $|$SNR$|$ cutoff of five. This can occur for features that are within the side lobes of strong features and are nominally removed from the catalogue to prevent the detection of spurious features \citepalias{FFtech}. \citetalias{Hopwood15} studied multiple observations of calibration sources for the \textit{Herschel} SPIRE FTS (see the following section) and marks a detected feature in the planetary nebula NGC 7027 at 802\,GHz that is not found by the FF in any FTS observations of the source since it occurs within the mask width of the bright CO(7--6) feature (rest frequency 806.65\,GHz) \citepalias{FFtech}. The low SNR search routine discovers this feature at 802.46\,GHz in multiple observations of NGC 7027 while looking for missing HCO$^+$ features. This feature is found at SNRs as high as 22 (observation ID 1342245075). 

\section{Results}
\subsection{Calibration Sources}
\label{sec:calSources}
%
\begin{table*}
\begingroup
\begin{center}
\newdimen\tblskip \tblskip=5pt
\caption{\label{tab:calibID}Summary of the line identification applied to the six line-rich single-pointing calibration sources. The number of high resolution observations are a selected subset of those described by \citetalias{Hopwood15} and are all point-source calibrated or a Hyper Processed Data Product (HPDP) \citepalias{Hopwood15}. The number of spectral features contained in SAFECAT (catalogued) is compared to the number of successful matches with the template (identified) and the lines added by our second pass of the FF (added) looking for low SNR features.}
\nointerlineskip
\small
%
\newdimen\digitwidth
\setbox0=\hbox{\rm 0}
\digitwidth=\wd0
\catcode`*=\active
\def*{\kern\digitwidth}
\newdimen\signwidth
\setbox0=\hbox{+}
\signwidth=\wd 0
\catcode`!=\active
\def!{\kern\signwidth}
%
\tabskip=2em plus 2em minus 2em
\halign to \hsize{\hfil#&\hfil*#\hfil& \hfil#\hfil& \hfil#\hfil&\hfil#\hfil&\hfil#\hfil&\hfil#\hfil&\hfil#\hfil& \hfil#*\hfil&#\hfil\cr
 &\multispan8\hrulefill& \cr
\noalign{\vspace{-8.0pt}}
 &\multispan8\hrulefill& \cr
& & & & & &\multispan3\hfil Spectral\,features \hfil& \cr
& Source & RA & Dec. & Type & HR Obs. & Catalogued & Identified & Added& \cr
\noalign{\vspace{-5.5pt}}
 &\multispan8\hrulefill& \cr
& NGC\,7027& 21:07:01.59& +42:14:10.2& Planetary nebula&             31& *\,832& 616& *89& \cr
& AFGL\,4106& 10:23:19.47& -59:32:04.9& Post-red supergiant& 29& *\,375& 338& *62& \cr
& CRL\,618$^a$& 04:45:53.64& +36:06:53.4& Protoplanetary nebula&     22& 1\,205& 832& 149& \cr
& CW\,Leo$^b$& 09:47:57.41& +13:16:43.6& Variable star&              *7& *\,705& 374& *56& \cr
& VY\,CMa& 07:22:58.33& -25:46:03.5& Red supergiant&                 *8& *\,620& 291& *60& \cr
& AFGL\,2688$^c$& 21:02:18.78& +36:41:41.2& Protoplanetary nebula& 17& *\,978& 602& *43& \cr
&\multispan8\hrulefill& \cr
&\multispan2\footnotesize{$^a$Also known as AFGL\,618.}& & & & & & & \cr 
&\multispan2\footnotesize{$^b$Also known as irc+10216.}& & & & & & & \cr 
&\multispan2\footnotesize{$^c$Also known as CRL\,2688.}& & & & & & & \cr 
\noalign{\vspace{-7.5pt}}
}
\end{center}
\endgroup
\vspace{-12pt}
\end{table*}

\begin{figure}
	\includegraphics[trim = 0mm 0mm 0mm 0mm, clip, width=\columnwidth]{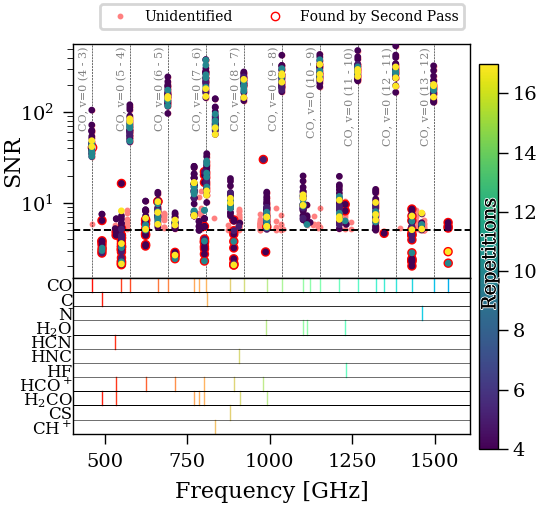}
    \vspace{-12pt}\caption{The line identification of features extracted by the FF and the inclusion of low SNR lines from 31 observations of the planetary nebula NGC 7027, a calibration source for the SPIRE spectrometer. Each point represents a spectral feature. Repetitions corresponds to the number of full scans (forward and reverse) of the FTS and is proportional to integration time. The dotted black line shows the cut-off that exists at SNR = 5.}
    \label{fig:NGC7027}
\end{figure}

The above line identification and low SNR search routines were first applied to single-pointing, line-rich, SPIRE FTS calibration sources (see \citetalias{Hopwood15}): AFGL 4106, CRL 618, NGC 7027, CW Leo, VY CMa, and AFGL 2688. These calibration sources were viewed multiple times throughout the SPIRE mission and make for excellent sources to test the line identification and uncovering of low SNR features. Each of these sources have confident radial velocity measurements in the FF based on their CO features \citepalias{FFredshift} and \citetalias{Hopwood15} provides measurements of some of the spectral lines observed by the SPIRE FTS for many of these sources. These published results can be compared against those from the line identification and low SNR search routines. Fig.\,\ref{fig:NGC7027} demonstrates the results of both the identification and low SNR search routines on 31 observations of NGC 7027. The low SNR search routine reveals a number of lines at low SNR that have been discovered in the source at $|\text{SNR}| \geq 5$ in some observations but fall below that threshold in others. Of the 31 observations of NGC 7027, 616 of 832 (74\%) lines contained in the FF catalogue have been successfully matched with the template and an additional 89 lines are found by the second pass of the FF looking for low SNR lines. The line identification results from matching with the template are presented in Tab.\,\ref{tab:calibID} with the number of lines found by the low SNR search routine for each of the calibration sources. 
NGC 7027 is confirmed to contain $^{13}$CO and HCO$^+$ emission features by \citetalias{Hopwood15} that are commonly excluded by the FF since they occur at low SNR in several observations. These account for the majority of the 89 features added by the low SNR search routine which also finds a combination of CH$^+$, p-H$_2$O, HF, and HNC features. All 10 in-band CO features are readily identified in observations of this source as is the [\ion{N}{II}] $^3$P$_1$--$^3$P$_0$ fine structure line. 

AFGL 4106 is a notably fainter source than other SPIRE FTS calibrators and observations of it contain a significant amount of emission from surrounding galactic cirrus (\citetalias{Hopwood15}, \citet{Molster1999, vanLoon99}). The [\ion{N}{II}] feature is easily identified in all observations of this source. All 10 of the in-band CO features are identified through template matching and are confirmed to be present by \citetalias{Hopwood15}. As in NGC 7027, the FF often misses low SNR $^{13}$CO features observed in the source which are successfully added to the catalogue by the low SNR search. The p-H$_2$O features are also identified in this source from the lines extracted by the FF, but they are not reported by \citetalias{Hopwood15}. There is evidence supporting the presence of H$_2$O ice in the dust envelope of AFGL 4106 \citep{Molster1999} which is likely linked to this water vapour emission observed by the SPIRE FTS \citep{VapourIceModel, waterAGBmodel}.

CRL 618 contains a significantly more diverse set of molecules than the previous two sources we have discussed. The presence of HCN, HNC, and H$_2$O features that are identified through template matching are also reported by \citetalias{Hopwood15}. In this source there are also incidences of the low SNR search finding $^{12}$CO features at a very high SNR that are not matched to the template during line identification. These features are indeed extracted by the FF but are not successfully identified since they fall just further than 0.3\,GHz tolerance from the corresponding template line. These features are included in the list of missing features that are searched for in the second pass of the FF and fit by the routine. The refitting of these features effectively widens the 0.3\,GHz matching window by the central frequency fitting constraints in the main loop of the FF (see \citetalias{FFtech}). This effective widening of the matching window can be seen for the $^{12}$CO $J=10$--$9$ transition in this source and for several other transitions in the calibration sources discussed in this paper.

\begin{figure}
	\includegraphics[trim = 5mm 4mm 0mm 2mm, clip, width=\columnwidth]{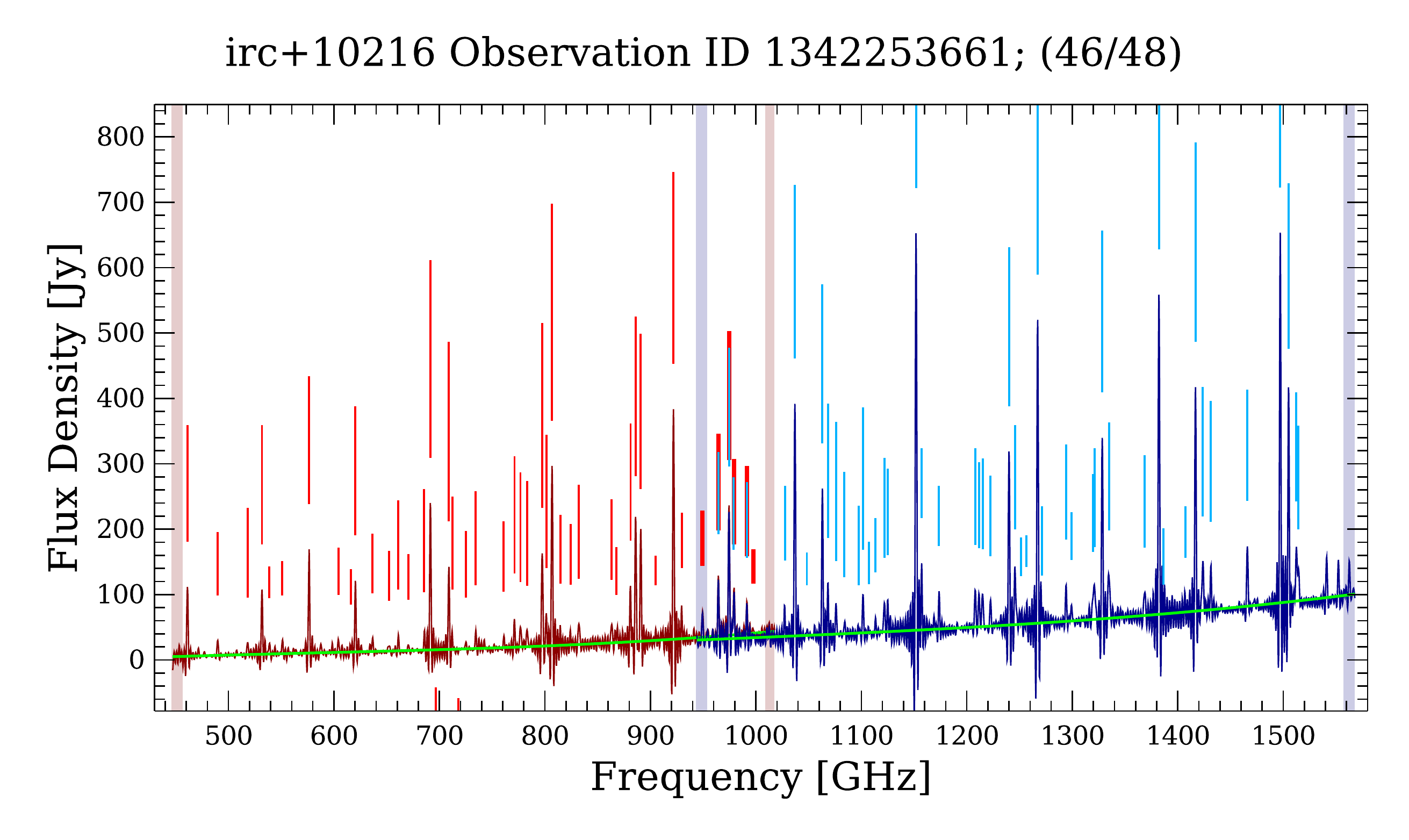}
	\includegraphics[trim = 5mm 7mm 0mm 2mm, clip, width=\columnwidth]{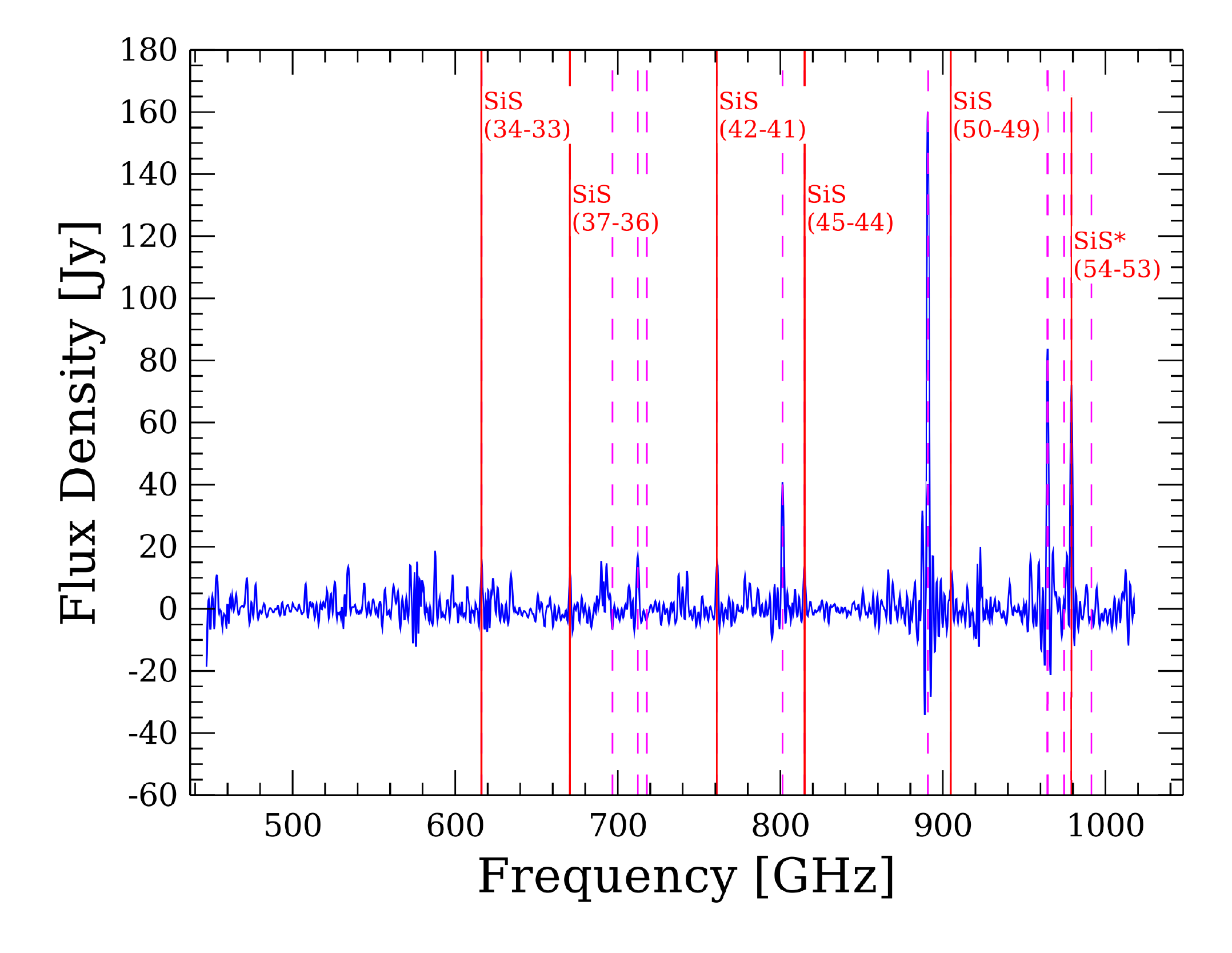}
	\includegraphics[trim = 5mm 7mm 0mm 2mm, clip, width=\columnwidth]{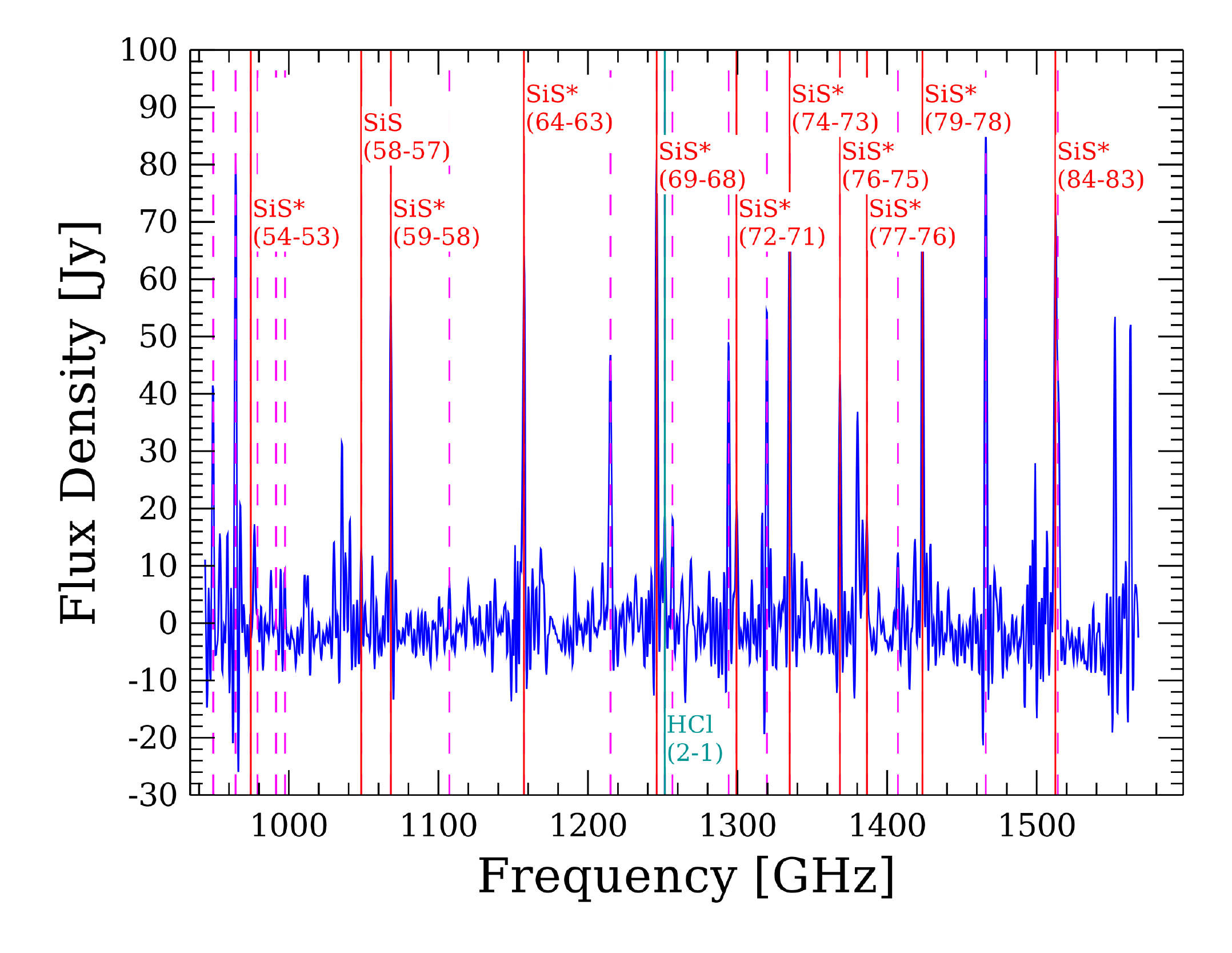}
    \vspace{-12pt}\caption{A single SPIRE observation of CW Leo (irc+10216) demonstrating the remaining spectral lines that have not been matched with the template for a single observation. The top panel is the FF postcard for the observation showing the spectrum from the SLWC3 (red) and SSWD4 (blue) central detectors. There are 94 lines extracted by the FF that are marked with vertical sticks above the spectrum. The bottom two panels show the continuum subtracted spectrum of the SLWC3 detector (middle) and SSWD4 detector (bottom) with the 67 lines that have been successfully matched with the template removed from the spectrum. Features extracted by the FF routine are marked by the dashed magenta lines. The marked SiS features are those identified by \citet{CWLeoSilicon}, those marked with an asterisk denote lines that are blended with other features.}
    \label{fig:CWLeoObs}
\end{figure}

In Tab.\,\ref{tab:calibID} we see that most of the FF extracted features from the nebular sources CRL 618 and NGC 7027 as well as the post-red supergiant AFGL 4106 are identified by the template but several lines found in the stellar sources VY CMa and CW Leo are not. CW Leo is a known source of HCl lines and SPIRE FTS observations of both CW Leo and VY CMa are known to contain several SiS lines \citep{CWLeoSilicon, CWLeoHCL, VYCMAsi}; these features are not part of the template since they are less ubiquitous than the other features that are included in the template. Fig.\,\ref{fig:CWLeoObs} shows the features in an observation of CW Leo that were not matched to the template. The FF fitted continuum has been subtracted and lines extracted by the FF that have been successfully matched to the template have been removed. Features extracted by the FF that have not been matched are marked by dashed magenta lines. Most of the lines that are not matched to the template are the rotational SiS features identified by \citet{CWLeoSilicon} (marked in red), we also note the HCl feature found at 1251\,GHz that is not included in the template (marked in cyan). SiS features which have been identified as blended features by \citet{CWLeoSilicon} are marked with an asterisk and can be as far as 2.52\,GHz from the fitted frequency of a FF extracted feature. Remaining spectral features that have not been labelled are likely some combination of contributions from SO$^+$, NH$_3$, and SiC molecules. There may also be contributions from molecular lines corresponding to higher vibrational states of water that are not included in the template line list. A large number of water lines have been observed in VY CMa by the \textit{Infrared Space Observatory} \citep{VYCMaWater} and a number of water transitions are also readily found in the 8 SPIRE FTS observations studied.

Of the spectral features from observations of AFGL 2688, 62\% are successfully identified through template matching. \citetalias{Hopwood15} confirm the presence of HCN, $^{12}$CO, and $^{13}$CO features that are commonly identified by the template matching and low SNR search. There is a high SNR feature detected across several observation of this source at 1414.77--1414.90\,GHz. This feature could possibly correspond to the H$_2$O, v=1 17(4, 14)--16(5, 11) transition at 1414.86\,GHz or a SiC feature at 1414.99\,GHz which are not included in the template. 

The full catalogue of SPIRE observations contains a variety of sources including nearby nebulae and stellar objects to high redshift galaxies \citep{wilson17}. Though we wish to provide as comprehensive a catalogue as possible, we do not intend for the line identification from matching spectral features to the template to replace a proper case by case analysis of each spectral observation. Many sources will contain spectral features that are not included in the template. The line identification should be considered a first look that will provide users with more information to determine which observations are of interest to them. The full line identification results for each calibration source broken down by atomic/molecular species is summarized in Tab.\,\ref{tab:linIDCalMol}.
%
\begin{table}
\begingroup
\begin{center}
\newdimen\tblskip \tblskip=5pt
\caption[Atomic/molecular species identified in the six calibration sources]{\label{tab:linIDCalMol} Atomic/molecular species identified in the six calibration sources.}
\nointerlineskip
\small
%
\newdimen\digitwidth
\setbox0=\hbox{\rm 0}
\digitwidth=\wd0
\catcode`*=\active
\def*{\kern\digitwidth}
\newdimen\signwidth
\setbox0=\hbox{+}
\signwidth=\wd 0
\catcode`!=\active
\def!{\kern\signwidth}
%
\tabskip=2em plus 2em minus 2em
\halign to \hsize{\hfil#&\hfil*#\hfil& \hfil#\hfil& \hfil#\hfil&\hfil#\hfil&\hfil#\hfil&\hfil#\hfil& \hfil#*\hfil&#\hfil\cr
 &\multispan7\hrulefill& \cr
\noalign{\vspace{-8.0pt}}
 &\multispan7\hrulefill& \cr
& & NGC& AFGL& CRL& CW& VY& AFGL& \cr
& & 7027& 4106& 618& Leo& CMa& 2688& \cr
\noalign{\vspace{-5.5pt}}
 &\multispan7\hrulefill& \cr
& CO&             310& 277& 208& *70& *65& 168& \cr
& $^{13}$CO&      287& 177& 198& *74& *28& 178& \cr
& C$^{17}$O&      **5& **0& *16& **3& **0& *24& \cr
& C$^{18}$O&      **0& **0& *38& **4& **8& *42& \cr
& \ion{N}{II}&    *12& *28& **0& **0& **0& **0& \cr
& \ion{C}{I}&     *39& **0& **0& **0& **0& **0& \cr
& H$_2$O &        *11& *11& 151& *25& 128& *85& \cr
& H$_2^{18}$O&    **6& **1& **4& **7& **8& **0& \cr
& H$_2$CO&        *32& *31& *54& *37& *40& *37& \cr
& HCO$^+$&        *80& **0& 181& **0& *16& **0& \cr
& H$^{13}$CO$^+$& **0& **0& **0& **2& **9& **0& \cr
& CH$^+$&         *31& **0& **0& **0& **0& **0& \cr
& CS&             **1& **1& **2& **0& **0& **6& \cr
& $^{13}$CS&      **0& **0& **0& **0& **0& **3& \cr
& HCN&            **1& **0& 254& 103& *70& 205& \cr
& H$^{13}$CN&     **0& **0& *17& **7& **8& 130& \cr
& HNC&            **1& **0& 130& *11& **0& **0& \cr
& HF&             **1& **0& **0& **0& **0& **0& \cr
& H$_2$S&         **0& **0& **3& **0& **0& **0& \cr
& SO&             **0& **0& **1& **0& **0& **1& \cr
& SiO&            **0& **0& **1& *56& 139& **0& \cr
& N$_2$H$^+$&     **0& **0& **0& **7& **0& **0& \cr
&\multispan7\hrulefill& \cr
\noalign{\vspace{-7.5pt}}
}
\end{center}
\endgroup
\vspace{-12pt}
\end{table}

\subsection{Full Catalogue}
Line identification was performed on all 1\,206 single-pointing sparse observations and 205 mapping observations. It is important to note that there is a degree of double counting in the number of pointing observations since each sparse observation has both a point source and extended source calibration. We follow the same convention as the FF catalogue for determining which calibration to use (or both) (see \citetalias{FFtech}). 

Only 283 of the lines extracted by the FF (1.4\%) from sparsely sampled single-pointing observations are without reliable velocity estimates. Of lines with velocity estimates, 77.3\% are successfully matched to the template and an additional 1\,516 features are found by the low SNR search. There are a large number of features in the 1261--1325\,GHz region that are unable to be matched with the template. Tab.\,\ref{tab:possibleTempGap} shows several spectral features in this region that are not included in the template. Many of these transitions are reported as possible matches to spectral features in SPIRE FTS observations of the calibration source VY CMa (see \citet{VYCMAsi}).
\begin{table}
	\centering
	\caption{Spectral features not included in the template that occur in the frequency region centred at 1288 GHz where many FF lines are not identified by the template for sparse single-pointing observations.}
	\label{tab:possibleTempGap}
	\begin{tabular}{ccc}
		\hline \hline
		\multicolumn{1}{c}{Molecule} & \multicolumn{1}{c}{Transition} & Rest frequency {[}GHz{]} \\
		\hline
		H$_2$O & 7(4,3)--6(5,2) & 1278.27 \\
		H$_2$O & 8(2,7)--7(3,4) & 1296.41 \\
		H$_2$O v2=3 & 6(5,2)--7(4,3) & 1286.83 \\ 
		HDO & 2(1,2)--1(0,1) & 1277.68 \\
		HDO & 2(2,0)--2(1,1) & 1291.64 \\
		HDO & 8(2,6)--8(2,7) & 1293.37 \\
		SiS & 71--70 & 1280.45 \\
		SiS & 72--71 & 1298.24 \\
		SO & 26(2,6)--25(2,6) & 1287.66  \\
		SO & 30(2,9)--29(2,8) & 1287.75  \\
		SO & 30(3,0)--29(2,9) & 1287.78  \\
		H$_2$S & 9(6,4)--9(5,5) & 1283.11 \\ 
		H$_2$S & 6(2,4)--6(1,5) & 1298.79  \\ 
		C$^{34}$S v=1 & 27--26 & 1289.17 \\
		CS & 26--25 & 1270.94 \\
		H$_2$CO & 17(2,15) -- 16(2,4) & 1269.48 \\
		HC$^{15}$N & 15--14 & 1289.73 \\
		\hline
	\end{tabular}
\vspace{-6pt}
\end{table}
Mapping observations provide a significantly larger data set of lines extracted by the FF than sparse observations and the majority of the observed sources are large extended emission regions. Mapping observation of the SPIRE FTS, unlike sparse observations, are kept in the HSA as hyperspectral cube products \citep{spire_handbook}. The FF treats the spectrum from each spaxel of a mapping observation separately \citepalias{FFtech}. Velocity estimates are only obtained by measurements of the $^{12}$CO rotational ladder or the [\ion{N}{II}] $^3$P$_1$--$^3$P$_0$ feature. These estimates are obtained by combining FF extracted spectral features from each spaxel in the SSW hyperspectral cube with FF features from the nearest SLW spaxel \citepalias{FFredshift}. We found that 14.1\% of lines from mapping observations are from spaxels that do not have corresponding a radial velocity reported by the FF. Mapping observations tend to be of sources with complex emission structure and dark spaxels that do not contain sufficient spectral information for a reliable velocity estimate are known to occur (see \citetalias{FFredshift}). 

Of the lines from spaxels with reliable velocity estimates, 95.2\% are successfully matched to the template by the line identification routine and an additional 370 lines are found by the low SNR search. There are a large number of unidentified features towards the edge of the SSW band (1531--1568\,GHz) where a series of H$_2$CO features that are not included in the template occur. Tab.\,\ref{tab:possibleTempGap2} shows these H$_2$CO features that occur in this region with some other lines that are also not included in the template. These transitions are possible candidates for the unidentified features. 
\begin{table}
	\centering
	\caption{Spectral features not included in the template that occur in the frequency region centred at 1549\,GHz where many FF lines from mapping observations are not identified by the template.}
	\label{tab:possibleTempGap2}
	\begin{tabular}{ccc}
		\hline \hline
		\multicolumn{1}{c}{Molecule} & \multicolumn{1}{c}{Transition} & Rest frequency {[}GHz{]} \\
		\hline
		H$_2$O & 6(3,3)--5(4,2) & 1541.97 \\
		H$_2$CO & 21(5,17) -- 20(5,16) & 1530.39\\
		H$_2$CO & 22(0,22) -- 21(0,21) & 1530.78\\
		H$_2$CO & 21(3,19) -- 20(3,18) & 1531.51\\
		H$_2$CO & 21(4,18) -- 20(4,17) & 1532.95\\
		H$_2$CO & 21(4,7) -- 20(4,16) & 1534.57\\
		H$_2$CO & 21(1,20) -- 20(1,19) & 1540.68\\
		H$_2$CO & 21(3,18) -- 20(3,17) & 1540.68\\
		HC$^{15}$N & 18 -- 17 & 1547.10\\
		H$^{13}$CN & 18 -- 17 & 1552.19\\ 
		HNC & 17--16 & 1539.33 \\
		\hline
	\end{tabular}
\vspace{-6pt}
\end{table}
\begin{figure*}
	\includegraphics[width=0.98\textwidth]{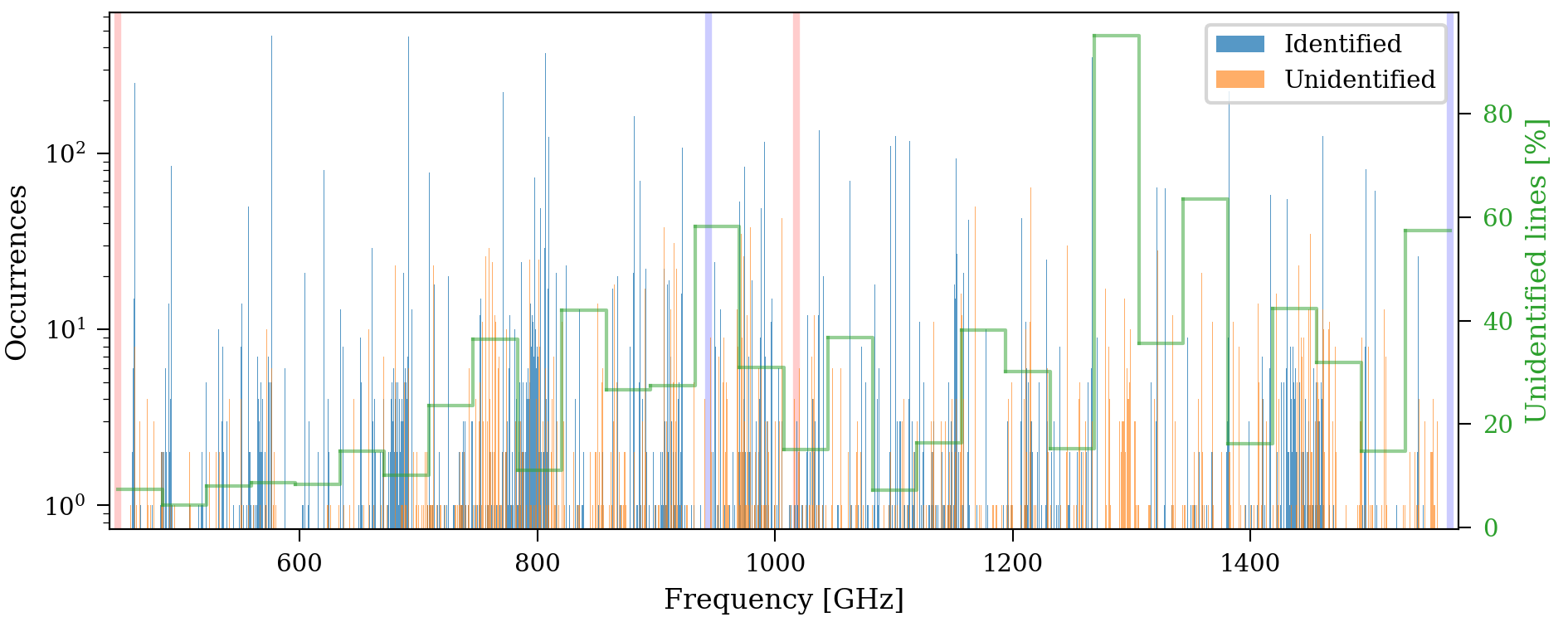}
	\includegraphics[width=0.98\textwidth]{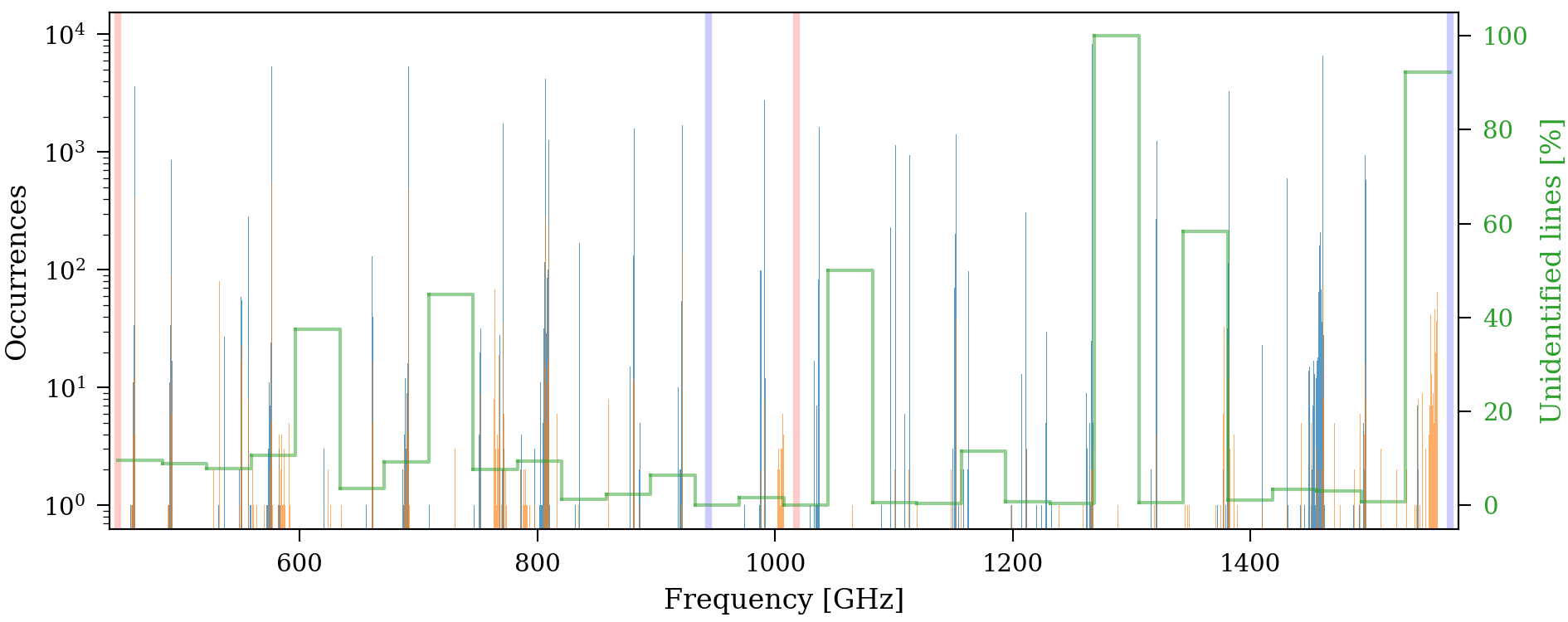}
	\includegraphics[width=0.98\textwidth]{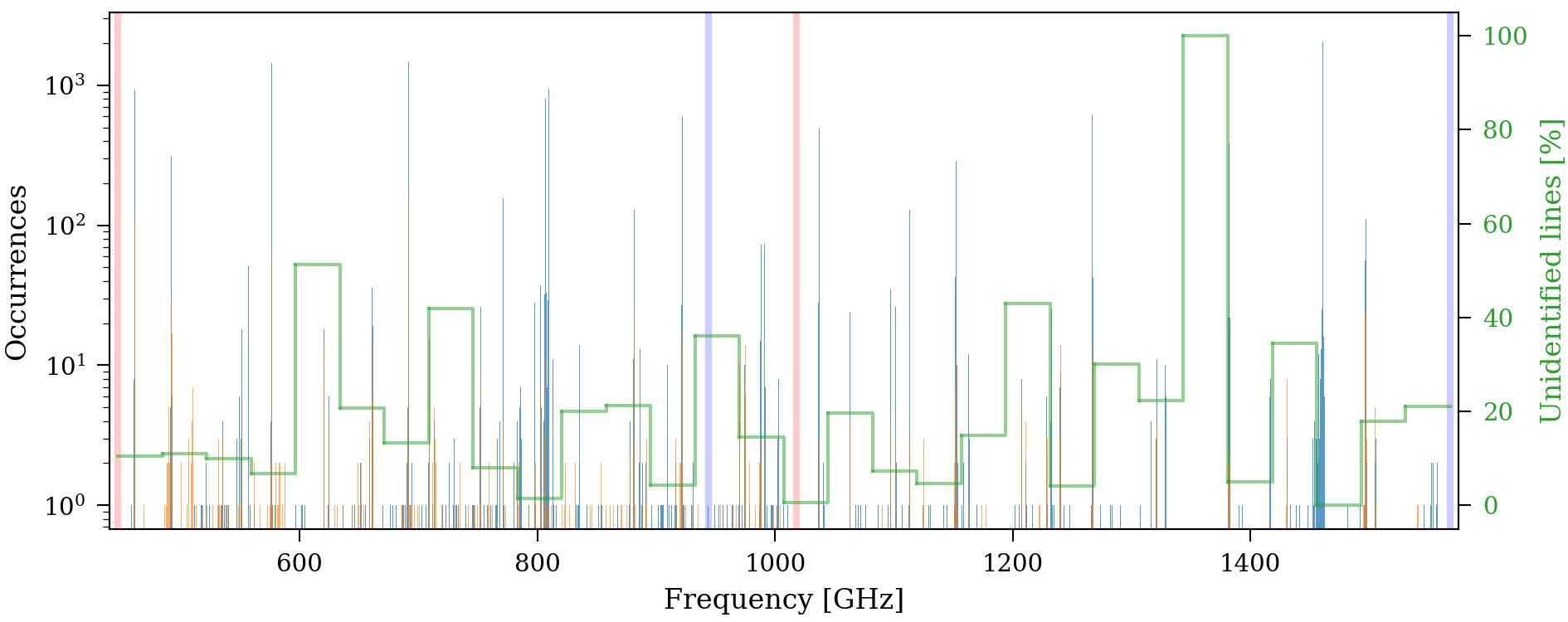}
    \caption{Histograms showing the number of lines identified from the FF. Identification results from the central detectors of sparsely sampled single-pointing observations, mapping observations, and off-axis detectors from sparsely sampled single-pointing observations (see \S\,\ref{sec:offAx}) are displayed in the top, middle, and bottom plot, respectively. Red and blue lines mark the extent of the SLW and SSW bands respectively, frequency bins match the template matching tolerance of 0.3 GHz. Larger green bins showing the number of unidentified features compared to the total number of features are also included to guide the eye.}
    \label{fig:lineHists}
\end{figure*}
The low SNR search routine, by design, is searching for features at low SNRs, where spurious detections of features are common and should be viewed with a degree of skepticism. A significant portion of features found by the low SNR search are considered a poor fit by the FF flagging criteria \citepalias{FFtech}. Of the lines found by the low SNR search, 75.9\% from sparse observation and 87.1\% from mapping observations are flagged as poor fits. Though caution should be used when considering these features, this routine fills a known weakness of the FF routine and aids in increasing the frequency spacing tolerance in the identification routine when an expected prominent feature is found by the FF more than 0.3\,GHz from its rest frequency. 

The results for both pointing and mapping observations are broken down into histograms over frequency in Fig.\,\ref{fig:lineHists}. Identified features are shown in the blue histogram while the FF features that remain unmatched to template lines are shown in the orange histogram. The bin widths of the histograms match the 0.3\,GHz template matching tolerance. The green bins in Fig.\,\ref{fig:lineHists} show the ratio between the number of lines extracted by the FF that remain unidentified by the template matching and the total number of lines from observations with reliable velocity estimates. This ratio is determined over larger frequency ranges than the bins of the histogram in order to guide the eye to regions of high or low success in line identification. From Fig.\,\ref{fig:lineHists}, we see that the line identification routine is very effective at matching lines from the CO ladder and the [\ion{N}{II}] $^3$P$_1$--$^3$P$_0$ 1461.1314\,GHz fine structure line which tend to be fairly bright in SPIRE observations. The rest frequency of all 10 in-band rotational transitions of CO are shown in Tab.\,\ref{tab:12COtrans}. Fig.\,\ref{fig:lineHists} also displays the results of the line identification routine for the off-axis detectors of sparsely sampled single-pointing observations which is the subject of the following section.

\begin{table}
	\centering
	\caption{The rotational transitions of $^{12}$CO. Transition numbers represent the total angular momentum quantum number, $J$.}
	\label{tab:12COtrans}
	\begin{tabular}{cccc}
		\hline \hline
		\multicolumn{2}{c}{SLW} & \multicolumn{2}{c}{SSW} \\
		Transition & Frequency {[}GHz{]} & Transition & Frequency {[}GHz{]} \\
		\hline
		4--3 & 461.041 & 9--8 & 1036.912 \\
		5--4 & 576.268 & 10--9 & 1151.985 \\
		6--5 & 691.473 & 11-10 & 1267.014 \\
		7--6 & 806.652 & 12--11 & 1381.995 \\
		8--7 & 921.800 & 13--12 & 1496.923 \\
		\hline
	\end{tabular}
\vspace{-12pt}
\end{table}

\vspace{-6pt}
\subsection{Off-Axis Identification}
The line identification routine successfully matches 23\,756 spectral features from the spectra provided by off-axis detectors (93.1\% of lines with velocity estimates) with their corresponding atomic/molecular transitions. These results are summarized in the bottom panel of Fig.\,\ref{fig:lineHists}. 

\begin{figure}
	\includegraphics[width=\columnwidth]{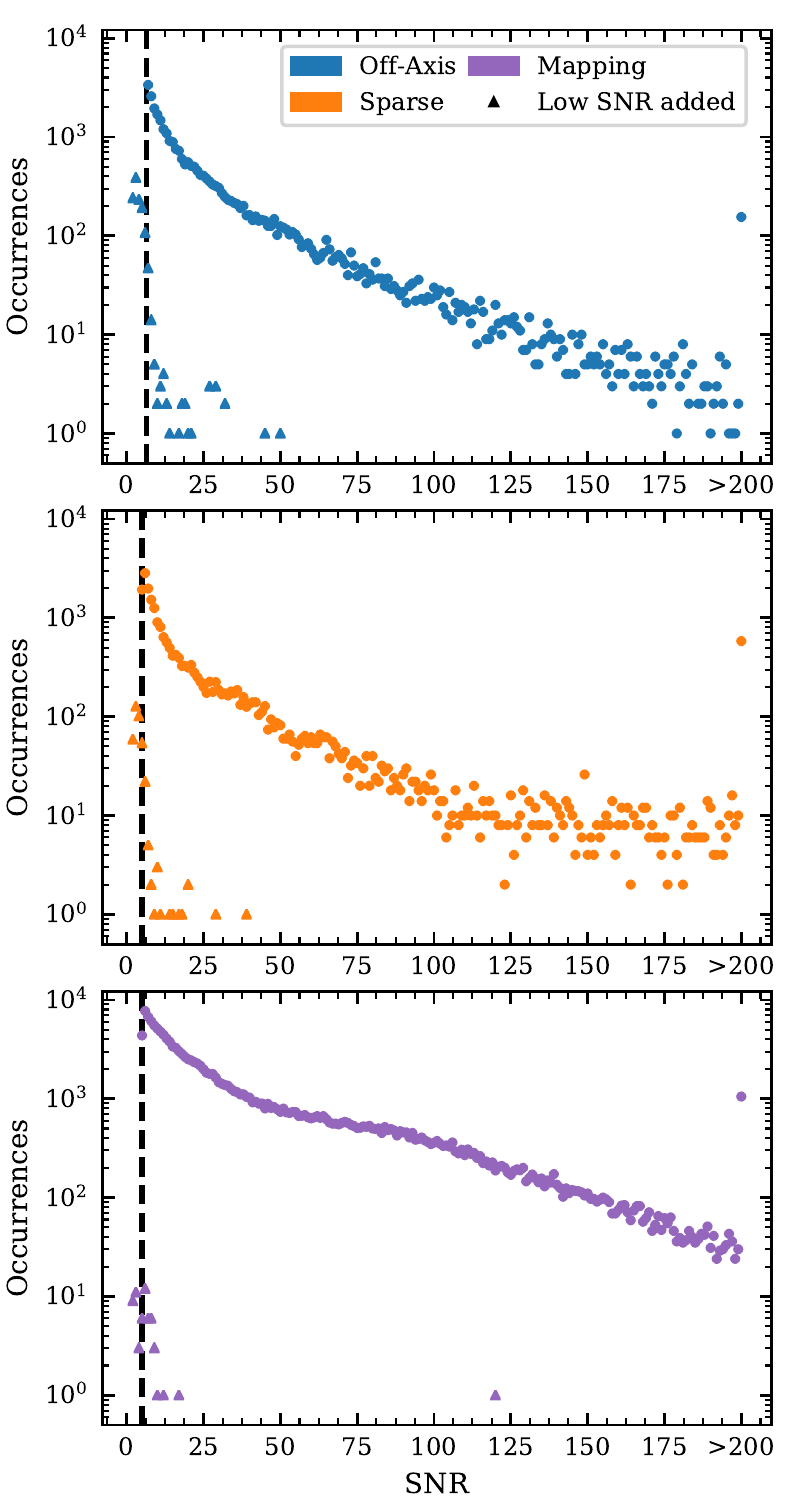}
	\vspace{-12pt}\caption{The SNR of lines extracted by the FF and the low SNR for off-axis detectors of sparse observations (top), central detectors of sparse observations (middle), and from mapping observations (bottom) are shown in blue, orange, and violet, respectively. Lines found by the FF employ round markers while those found by the low SNR search routine through line identification use triangular markers (see \S\,\ref{sec:lowSNR}) . The $|$SNR$|$ cutoff in each case is shown by the dotted black line. The axes in each case have been set to the same scale to allow comparison.} 
	\label{fig:offAxFullHist}
\end{figure}

The SNR results for off-axis detectors are compared to the rest of the catalogue in Fig.\,\ref{fig:offAxFullHist}. The SNR histogram of the features from off-axis pixels is shown on the top plot in blue while the results from the central detectors of sparse observations and the results from mapping observations are shown in orange on the middle plot and in violet on the final plot, respectively. The $|$SNR$|$ cut-off for each case is shown by the dashed black line and axis ranges are set to be equal for to allow comparison between each set of extracted features. We found that the spectra from off-axis detectors tend to be dimmer and contain lower SNR features than spectra from the central detectors (See Figs.\,\ref{fig:offAxIntensHist} and \ref{fig:offAxFullHist}), thus spurious detections of features are more common. A $|$SNR$| \geq 6.5$ cutoff was chosen for features extracted by the FF for off-axis pixels as opposed to the nominal cutoff of $|$SNR$| \geq 5$ for the central detectors in sparse observations and for mapping observations. Features discovered by the low SNR search routine are shown by triangular shaped markers. A total of 1\,234 features from off-axis detectors in sparse observations have been added to the catalogue by this routine. 

\vspace{-6pt}
\section{Conclusions}
\begin{figure}
	\includegraphics[trim = 2mm 2mm 1mm 1mm, clip, width=\columnwidth]{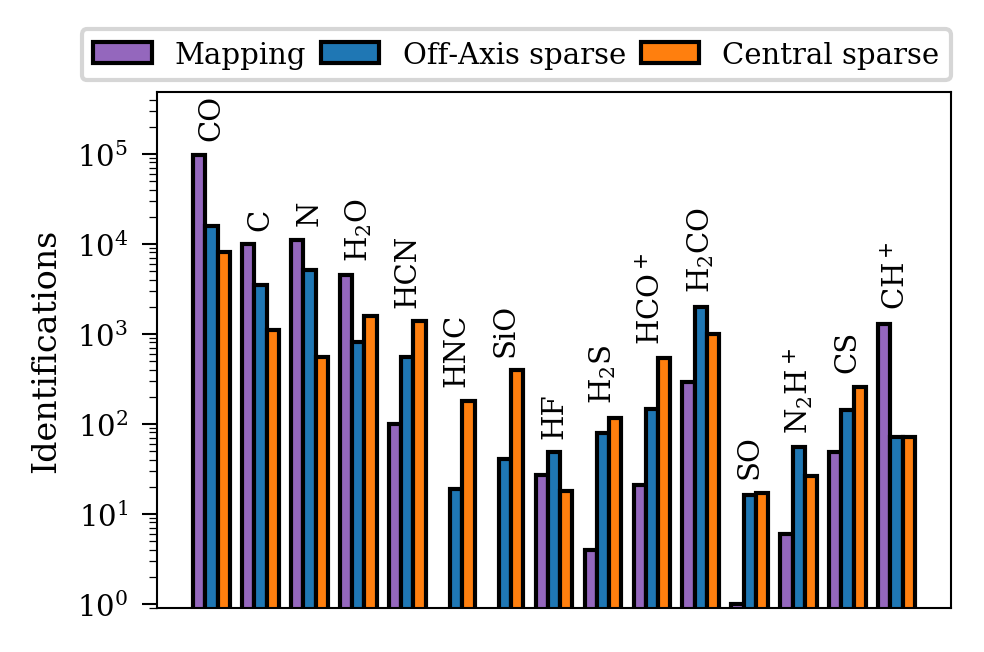}
	\vspace{-14pt}\caption{The line identification of all 15 atomic/molecular species contained in the template for mapping and sparse SPIRE observations.} 
	\label{fig:fullIDSpecies}
	\vspace{-12pt}
\end{figure}
The extensions of the FF to include spectral observations from off-axis detectors in sparsely sampled single-pointing observations is detailed. An additional 30\,720 spectral features with a $|$SNR$|$ greater than 6.5 are aggregated into the SPIRE automated Feature Extraction Catalogue (SAFECAT) by the inclusion of the analysis and extraction of features from the off-axis detector spectra. We developed a set of postcards to highlight and summarize results from these off-axis detectors in each observation that are of a similar format to the mapping postcards described by \citetalias{FFtech} and are available from the ESA Herschel legacy area.\textsuperscript{\ref{ff_legacy}}.

We have also extended the automated SPIRE Feature Finder (FF) routine \citepalias{FFtech} to identify the atomic or molecular transition that corresponds to the features extracted from HR observations of the  SPIRE FTS. This identification of atomic/molecular transitions is done by matching the spectral features to a template of 307 atomic fine-structure and molecular features that are commonly found in SPIRE FTS observations. 
Line identification is reliant on the radial velocity estimates included in the catalogue by \citetalias{FFredshift} and is able to successfully match 77.3\%, 93.1\%, and 95.2\% of spectral features found by the FF from central detectors of sparse single-pointing observations, off-axis detectors, and mapping observations, respectively, to their atomic/molecular transitions. We find this line identification to be particularly effective at identifying the CO rotational ladder from $J=4$--$3$ to $J=13$--$12$ and the [\ion{N}{II}] $^3$P$_1$--$^3$P$_0$ transition in SPIRE FTS observations. The identification of all 15 atomic/molecular species in SPIRE FTS spectra from mapping, off-axis detectors in sparse observations, and central detectors in sparse observations is shown in Fig.\,\ref{fig:fullIDSpecies}. Note that isotopologues and ionization states have all been grouped together for each species.

The information provided by the line identification is also used to uncover low SNR features that fall below the $|$SNR$|\geq5$ cutoff that exists nominally in the FF \citepalias{FFtech}. This low SNR search routine adds an additional 1\,516 for the central detectors of sparsely sampled single-pointing observations and 1\,234 for the off-axis detectors. An additional 370 lines are added by this routine from mapping observations of the SPIRE FTS. 

\vspace{-12pt}
\section*{Acknowledgements}
\emph{Herschel} is an ESA space observatory with science instruments provided by European-led Principal Investigator consortia and with important participation from NASA\@. This research acknowledges support from ESA, CSA, CRC, CMC, and NSERC\@.

This research has made use of the NASA/IPAC Infrared Science Archive, which is funded by the National Aeronautics and Space Administration and operated by the California Institute of Technology.

This research has made use of the \textsc{SciPy} (\url{www.scipy.org}) and \textsc{Astropy} (\url{www.astropy.org}) Python packages. Table formatting in this paper followed the {\it Planck} Style Guide \citep{PlanckStyle}.
\nocite{2020SciPy-NMeth, astropy:2013, astropy:2018}\\

\section*{Data Availability}
\vspace{-3pt}
The \textit{Herschel} SPIRE Spectral Feature Catalogue has been assigned an ESA Digital Object Identifier (DOI) and is available at: \href{https://doi.org/10.5270/esa-lysf2yi}{doi.org/10.5270/esa-lysf2yi}. The FF code and all FF products are publicly available via the \textit{Herschel} Science Archive.



\nocite{chrisThesis}
\bibliographystyle{mnras}
\bibliography{LineID_paper_FinalSubmission} 



\vspace{-12pt}
\appendix

\section{Line Matching Template}
A note on the \textbf{Quantum Numbers} presented in Tab.\,\ref{tab:catalogue}. For molecules where the vibrational band (denoted `v') is not explicitly stated, it is zero. Sometimes the vibrational band is expressed over a range (e.g. CS, v=0-4). J in all cases refers to the total angular momentum quantum number (excluding nuclear spin) and transitions from primed quantum numbers to unprimed. Quantum numbers of the format, J$^\prime$(K$^\prime$$_a$,K$^\prime$$_c$) -- J(K$_a$,K$_c$) such as for water molecules are arranged such that J is the total angular momentum quantum number while K$_a$ and K$_c$ are the limiting prolate and limiting oblate quantum number for the projection of total angular momentum along the axis of symmetry. Transitions marked by the transition of two numbers, e.g., J$^\prime$--J indicate a purely rotational state. The quantum number format J$^\prime$ F$^\prime$ -- J F indicates J as the total angular momentum quantum number and F indicates the total angular momentum quantum number \textit{including} nuclear spin.

\begin{table*}
\begingroup
\begin{center}
\newdimen\tblskip \tblskip=5pt
	\caption{The collection of 307 molecular, atomic, and fine-structure lines contained in the template that is matched to the lines catalogued by the FF routine. The full template was compiled from a list of commonly occurring spectral features in SPIRE observations. For full details see \citet{templateCassis}.}
	\label{tab:catalogue}
\nointerlineskip
\small
%
\newdimen\digitwidth
\setbox0=\hbox{\rm 0}
\digitwidth=\wd0
\catcode`*=\active
\def*{\kern\digitwidth}
\newdimen\signwidth
\setbox0=\hbox{+}
\signwidth=\wd 0
\catcode`!=\active
\def!{\kern\signwidth}
%
\tabskip=2em plus 2em minus 2em
\halign to \hsize{\hfil#& *#\hfil& \hfil#\hfil& \hfil#\hfil& *#\hfil&\hfil#\hfil&\hfil#\hfil& *#\hfil&\hfil#\hfil& \hfil#*\hfil&#\hfil\cr
 &\multispan9\hrulefill& \cr
\noalign{\vspace{-8.0pt}}
 &\multispan9\hrulefill& \cr
& Atom/& Quantum& $\nu_0$& Molecule& Quantum& $\nu_0$& Molecule& Quantum& $\nu_0$& \cr
& Molecule& Numbers& [GHz]& & Numbers& [GHz]& & Numbers& [GHz]& \cr
\noalign{\vspace{-5.5pt}}
 &\multispan9\hrulefill& \cr
& $[$\ion{C}{I}$]$& $^3$P$_1$--$^3$P$_0$&          *\,492.1607& H$^{13}$CN, v=0 & 11 -- 10 &     *\,949.3011& C$^{18}$O & 5 -- 4 & *\,548.8310& \cr 
& $[$\ion{C}{I}$]$& $^3$P$_2$--$^3$P$_1$&          *\,809.3420& H$^{13}$CN, v=0 & 12 -- 11 &     1\,035.5096& C$^{18}$O & 6 -- 5 & *\,658.5533& \cr 
& $[$\ion{C}{II}$]$& $^3$P$_{3/2}$--$^3$P$_{1/2}$& 1\,900.5369& H$^{13}$CN, v=0 & 13 -- 12 &     1\,121.6941& C$^{18}$O & 7 -- 6 & *\,768.2516& \cr 
& $[$\ion{N}{II}$]$& $^3$P$_1$--$^3$P$_0$&         1\,461.1314& H$^{13}$CN, v=0 & 14 -- 13 &     1\,207.8529& C$^{18}$O & 8 -- 7 & *\,877.9219& \cr 
& CH$^+$& 1 -- 0&                                  *\,835.1375& H$^{13}$CN, v=0 & 19 -- 18 &     1\,638.1891& C$^{18}$O & 9 -- 8 & *\,987.5604& \cr 
& CH$^+$& 2 -- 1&                                  1\,669.2813& H$^{13}$CN, v=0 & J = 20 -- 19 & 1\,724.1510& C$^{18}$O & 10 -- 9 & 1\,097.1629& \cr 
& p-H$_2$O& 6(2,4) -- 7(1,7)&                      *\,488.4911& H$^{13}$CN, v=0 & 21 -- 20 &     1\,810.0731 & C$^{18}$O & 11 -- 10 & 1\,206.7255& \cr 
& p-H$_2$O& 2(1,1) -- 2(0,2)&     *\,752.0332& H$^{13}$CN, v=0 & 22 -- 21 &     1\,895.9534 & C$^{18}$O & 12 -- 11 & 1\,316.2441& \cr 
& p-H$_2$O& 4(2,2) -- 3(3,1)&     *\,916.1716& H$^{13}$CN, v2=1 & 6 -1 -- 5 1 & *\,517.8180 & C$^{18}$O & 13 -- 12 & 1\,425.7149& \cr 
& p-H$_2$O& 5(2,4) -- 4(3,1)&     *\,970.3150& H$^{13}$CN, v2=1 & 6 1 -- 5 -1 & *\,520.3939 & $^{13}$C$^{17}$O & 4 1.5 -- 3 2.5 & *\,429.1170& \cr 
& p-H$_2$O& 2(0,2) -- 1(1,1)&     *\,987.9267& HNC, v=0 & J = 6 -- 5 &   *\,543.8976 & $^{13}$C$^{17}$O & 5 2.5 -- 4 3.5 & *\,536.3678& \cr 
& p-H$_2$O& 1(1,1) -- 0(0,0)&     1\,113.3430& HNC, v=0 & J = 7 -- 6 &   *\,634.5108 & $^{13}$C$^{17}$O & 12 9.5 -- 11 10.5 & 1\,286.3764& \cr 
& p-H$_2$O& 4(2,2) -- 4(1,3)&     1\,207.6388& HNC, v=0 & J = 8 -- 7 &   *\,725.1073 & $^{13}$C$^{17}$O & 13 10.5 -- 12 11.5 & 1\,393.3680& \cr 
& p-H$_2$O& 2(2,0) -- 2(1,1)&     1\,228.7887& HNC, v=0 & J = 9 -- 8 &   *\,815.6847 & $^{13}$C$^{18}$O & 5 4.5 -- 4 4.5 & *\,523.4842& \cr 
& o-H$_2$O& 1(1,0) -- 1(0,1)&     *\,556.9360& HNC, v=0 & J = 10 -- 9 &  *\,906.2405 & $^{13}$C$^{18}$O & 6 5.5 -- 5 5.5 & *\,628.1411& \cr 
& o-H$_2$O& 5(3,2) -- 4(4,1)&     *\,620.7008& HNC, v=0 & J = 11 -- 10 & *\,996.7723 & N$_2$H$^+$, v=0 & 6 -- 5 & *\,558.9665& \cr 
& o-H$_2$O& 3(1,2) -- 3(0,3)&     1\,097.3647& HNC, v=0 & J = 13 -- 12 & 1\,177.7547 & N$_2$H$^+$, v=0 & 7 -- 6 & *\,652.0956& \cr 
& o-H$_2$O& 3(1,2) -- 2(2,1)&     1\,153.1268& CO, v=0 & 4 -- 3 &     *\,461.0408 & N$_2$H$^+$, v=0 & 8 -- 7 & *\,745.2099& \cr 
& o-H$_2$O& 3(2,1) -- 3(1,2)&     1\,162.9116& CO, v=0 & 5 -- 4 &     *\,576.2679 & N$_2$H$^+$, v=0 & 9 -- 8 & *\,838.3073& \cr 
& o-H$_2$O& 5(2,3) -- 5(1,4)&     1\,410.7319& CO, v=0 & 6 -- 5 &     *\,691.4731 & N$_2$H$^+$, v=0 & 10 -- 9 & *\,931.3857& \cr 
& o-H$_2$O& 2(2,1) -- 2(1,2)&     1\,661.0076& CO, v=0 & 7 -- 6 &     *\,806.6518 & N$_2$H$^+$, v=0 & 11 -- 10 & 1\,024.4430& \cr 
& o-H$_2$O& 2(1,2) -- 1(0,1)&     1\,669.9047& CO, v=0 & 8 -- 7 &     *\,921.7997 & N$_2$H$^+$, v=0 & 12 -- 11 & 1\,117.4771& \cr 
& o-H$_2$O& 3(0,3) -- 2(1,2)&     1\,716.7696& CO, v=0 & 9 -- 8 &     1\,036.9124 & HCO$^+$, v=0 & 6 -- 5 & *\,535.0616& \cr 
& HDO& 1 1 1 -- 0 0 0&            *\,893.6387& CO, v=0 & 10 -- 9 &    1\,151.9855 & HCO$^+$, v=0 & 7 -- 6 & *\,624.2084& \cr 
& H$_2^{18}$O& 1(1,0) -- 1(0,1)&  *\,547.6764& CO, v=0 & 11 -- 10 &   1\,267.0145 & HCO$^+$, v=0 & 8 -- 7 & *\,713.3412& \cr 
& H$_2^{18}$O& 1(1,1) -- 0(0,0) & 1\,101.6983& CO, v=0 & 12 -- 11 &   1\,381.9951 & HCO$^+$, v=0 & 9 -- 8 & *\,802.4582& \cr 
& HF& 1 -- 0&                     1\,232.4762& CO, v=0& 13 -- 12 &    1\,496.9229 & HCO$^+$, v=0 & 10 -- 9 & *\,891.5573& \cr 
& HCN, v=0& 6 -- 5&               *\,531.7164& CO, v=0 & 14 -- 13 &   1\,611.7935 & HCO$^+$, v=0 & 11 -- 10 & *\,980.6365& \cr 
& HCN, v=0& 7 -- 6 &              *\,620.3040& CO, v=0 & 15 -- 14 &   1\,726.6025 & HCO$^+$, v=0 & 12 -- 11 & 1\,069.6939& \cr 
& HCN, v=0& 8 -- 7 &              *\,708.8770& CO, v=0 & 16 -- 15 &   1\,841.3455 & HCO$^+$, v=0 & 13 -- 12 & 1\,158.7272& \cr 
& HCN, v=0& 9 -- 8 &              *\,797.4333& $^{13}$CO & 4 -- 3 &   *\,440.7652 & HCO$^+$, v=0 & 14 -- 13 & 1\,247.7350& \cr 
& HCN, v=0& 10 -- 9 &             *\,885.9707& $^{13}$CO & 5 -- 4 &   *\,550.9263 & H$^{13}$CO$^+$ & 6 -- 5 & *\,520.4599& \cr 
& HCN, v=0& 11 -- 10 &            *\,974.4872& $^{13}$CO & 6 -- 5 &   *\,661.0672 & H$^{13}$CO$^+$ & 7 -- 6 & *\,607.1746& \cr 
& HCN, v=0& 12 -- 11 &            1\,062.9807& $^{13}$CO & 7 -- 6 &   *\,771.1841 & H$^{13}$CO$^+$ & 8 -- 7 & *\,693.8763& \cr 
& HCN, v=0& 13 -- 12 &            1\,151.4491& $^{13}$CO & 8 -- 7 &   *\,881.2728 & H$^{13}$CO$^+$ & 9 -- 8 & *\,780.5628& \cr 
& HCN, v=0& 14 -- 13 &            1\,239.8902& $^{13}$CO & 9 -- 8 &   *\,991.3293 & H$^{13}$CO$^+$ & 10 -- 9 & *\,867.2324& \cr 
& HCN, v=0& J = 15 -- 14&         1\,328.3084& $^{13}$CO & 10 -- 9 &  1\,101.3496 & HC$^{18}$O$^+$ & 6 -- 5 & *\,510.9096& \cr 
& HCN, v=0& J = 16 -- 15 &        1\,416.6830& $^{13}$CO & 11 -- 10 & 1\,211.3296 & p-H$_2$CO & 6(2,4) -- 7(0,7) & *\,485.2276& \cr 
& HCN, v=0& 17 -- 16 &            1\,505.0299& $^{13}$CO & 12 -- 11 & 1\,321.2655 & p-H$_2$CO & 7(0,7) -- 6(0,6) & *\,505.8337& \cr 
& HCN, v=0& 18 -- 17 &            1\,593.3415& $^{13}$CO & 13 -- 12 & 1\,431.1530 & p-H$_2$CO & 7(2,6) -- 6(2,5) & *\,509.1462& \cr 
& HCN, v=0& 19 -- 18 &            1\,681.6155& $^{13}$CO & 14 -- 13 & 1\,540.9883 & p-H$_2$CO & 7(4,4) -- 6(4,3) & *\,509.8296& \cr 
& HCN, v=0& 20 -- 19 &            1\,769.8498& $^{13}$CO & 15 -- 14 & 1\,650.7673 & p-H$_2$CO & 7(4,3) -- 6(4,2) & *\,509.8302& \cr 
& HCN, v=0& 21 -- 20 &            1\,858.0422& $^{13}$CO & 16 -- 15 & 1\,760.4860 & p-H$_2$CO & 7(2,5) -- 6(2,4) & *\,513.0763& \cr 
& HCN, v2=1& 6 -1 -- 5 1&         *\,531.6484& $^{13}$CO & 17 -- 16 & 1\,870.1404 & p-H$_2$CO & 8(0,8) -- 7(0,7) & *\,576.7083& \cr 
& HCN, v2=1& 6 1 -- 5 -1&         *\,534.3399& C$^{17}$O & 4 1.5 -- 3 2.5 &    *\,449.3947 & p-H$_2$CO & 8(2,7) -- 7(2,6) & *\,581.6118& \cr 
& HCN, v2=3& 6 -1 -- 5 1&         *\,532.9930& C$^{17}$O & 5 -- 4 &            *\,561.7128 & p-H$_2$CO & 8(6,3) -- 7(6,2) & *\,582.0708& \cr 
& HCN, v2=3& 6 -3 -- 5 3&         *\,535.2362& C$^{17}$O & 6 -- 5 &            *\,674.0094 & p-H$_2$CO & 8(4,5) -- 7(4,4) & *\,582.7229& \cr 
& HCN, v2=3& 6 1 -- 5 -1&         *\,538.5339& C$^{17}$O & 7 -- 6 &            *\,786.2808 & p-H$_2$CO & 8(4,4) -- 7(4,3) & *\,582.7242& \cr 
& HCN, v3=1& 6 -- 5&              *\,528.0899& C$^{17}$O & 8 -- 7 &            *\,898.5230 & p-H$_2$CO & 8(2,6) -- 7(2,5) & *\,587.4537& \cr 
& H$^{13}$CN, v=0& 6 -- 5&        *\,517.9698& C$^{17}$O & 9 6.5 -- 8 7.5 &    1\,010.7311 & p-H$_2$CO & 9(0,9) -- 8(0,8) & *\,647.0818& \cr 
& H$^{13}$CN, v=0& 7 -- 6&        *\,604.2679& C$^{17}$O & 10 11.5 -- 9 10.5 & 1\,122.9025 & p-H$_2$CO & 9(2,8) -- 8(2,7) & *\,653.9701& \cr 
& H$^{13}$CN, v=0& 8 -- 7&        *\,690.5521& C$^{17}$O & 11 8.5 -- 10 9.5 &  1\,235.0315 & p-H$_2$CO & 9(6,4) -- 8(6,3) & *\,654.8382& \cr 
& H$^{13}$CN, v=0& 9 -- 8&        *\,776.8203& C$^{17}$O & 12 9.5 -- 11 10.5 & 1\,347.1148 & p-H$_2$CO & 9(4,6) -- 8(4,5) & *\,655.6399& \cr 
& H$^{13}$CN, v=0& 10 -- 9&       *\,863.0706& C$^{18}$O & 4 -- 3 &            *\,439.0888 & p-H$_2$CO & 9(4,5) -- 8(4,4) & *\,655.6437& \cr 
&\multispan9\hrulefill& \cr
\noalign{\vspace{-7.5pt}}
}
\end{center}
\endgroup
\vspace{-12pt}
\end{table*}

\begin{table*}
	\centering
	\contcaption{}
	\label{tab:catalogueCont1}
\begingroup
\begin{center}
\newdimen\tblskip \tblskip=5pt
\nointerlineskip
\small
%
\newdimen\digitwidth
\setbox0=\hbox{\rm 0}
\digitwidth=\wd0
\catcode`*=\active
\def*{\kern\digitwidth}
\newdimen\signwidth
\setbox0=\hbox{+}
\signwidth=\wd 0
\catcode`!=\active
\def!{\kern\signwidth}
%
\tabskip=2em plus 2em minus 2em
\halign to \hsize{\hfil#& *#\hfil& \hfil#\hfil& \hfil#\hfil& *#\hfil&\hfil#\hfil&\hfil#\hfil& *#\hfil&\hfil#\hfil& \hfil#*\hfil&#\hfil\cr
 &\multispan9\hrulefill& \cr
\noalign{\vspace{-8.0pt}}
 &\multispan9\hrulefill& \cr
& & Quantum& $\nu_0$& Molecule& Quantum& $\nu_0$& Molecule& Quantum& $\nu_0$& \cr
& Molecule& Numbers& [GHz]& & Numbers& [GHz]& & Numbers& [GHz]& \cr
\noalign{\vspace{-5.5pt}}
 &\multispan9\hrulefill& \cr

& p-H$_2$CO & 9(2,7) -- 8(2,6) &     *\,662.2091& o-H$_2$CO & 13(3,11) -- 12(3,10) & *\,948.4538& $^{13}$CS, v=0,1 & 19 0 -- 18 0 & *\,877.7272& \cr 
& p-H$_2$CO & 10(0,10) -- 9(0,9) &   *\,716.9384& o-H$_2$CO & 13(1,12) -- 12(1,11) & *\,970.1992& $^{13}$CS, v=0,1 & 20 0 -- 19 0 & *\,923.8120& \cr 
& p-H$_2$CO & 10(2,9) -- 9(2,8) &    *\,726.2083& o-H$_2$CO & 14(1,14) -- 13(1,13) & *\,978.5924& $^{13}$CS, v=0,1 & 21 0 -- 20 0 & *\,969.8797& \cr 
& p-H$_2$CO & 10(4,7) -- 9(4,6) &    *\,728.5831& o-H$_2$CO & 14(3,12) -- 13(3,11) & \,1021.5331& $^{13}$CS, v=0,1 & 22 0 -- 21 0 & 1\,015.9294& \cr 
& p-H$_2$CO & 10(4,6) -- 9(4,5) &    *\,728.5916& o-H$_2$CO & 14(3,11) -- 13(3,10) & 1\,024.2886& SiO, v=0-10 & 12 1 -- 11 1 & *\,517.2597& \cr 
& p-H$_2$CO & 10(2,8) -- 9(2,7) &    *\,737.3427& o-H$_2$CO & 14(1,13) -- 13(1,12) & 1\,043.2229& SiO, v=0-10 & 12 0 -- 11 0 & *\,520.8811& \cr 
& p-H$_2$CO & 11(0,11) -- 10(0,10) & *\,786.2849& o-H$_2$CO & 15(3,13) -- 14(3,12) & 1\,094.5899& SiO, v=0-10 & 13 0 -- 12 0 & *\,564.2490& \cr 
& p-H$_2$CO & 11(2,10) -- 10(2,9) &  *\,798.3134& o-H$_2$CO & 16(3,14) -- 15(3,13) & 1\,167.6085& SiO, v=0-10 & 14 0 -- 13 0 & *\,607.6076& \cr 
& p-H$_2$CO & 11(2,9) -- 10(2,8) &   *\,812.8314& p-H$_2$S & 5(3,3) -- 6(0,6) & *\,493.0849& SiO, v=0-10 & 15 0 -- 14 0 & *\,650.9561& \cr 
& p-H$_2$CO & 12(0,12) -- 11(0,11) & *\,855.1513& p-H$_2$S & 3(3,1) -- 3(2,2) & *\,568.0506& SiO, v=0-10 & 16 0 -- 15 0 & *\,694.2939& \cr 
& p-H$_2$CO & 12(2,11) -- 11(2,10) & *\,870.2735& p-H$_2$S & 4(2,2) -- 4(1,3) & *\,665.3937& SiO, v=0-10 & 17 0 -- 16 0 & *\,737.6202& \cr 
& p-H$_2$CO & 12(2,10) -- 11(2,9) &  *\,888.6291& p-H$_2$S & 2(0,2) -- 1(1,1) & *\,687.3034& SiO, v=0-10 & 18 0 -- 17 0 & *\,780.9346& \cr 
& p-H$_2$CO & 13(0,13) -- 12(0,12) & *\,923.5878& p-H$_2$S & 3(2,2) -- 3(1,3) & *\,747.3019& SiO, v=0-10 & 19 0 -- 18 0 & *\,824.2359& \cr 
& p-H$_2$CO & 13(2,12) -- 12(2,11) & *\,942.0766& p-H$_2$S & 3(1,3) -- 2(0,2) & 1\,002.7787& SiO, v=0-10 & 20 0 -- 19 0 & *\,867.5236& \cr 
& p-H$_2$CO & 13(2,11) -- 12(2,10) & *\,964.6681& p-H$_2$S & 4(1,3) -- 4(0,4) & 1\,018.3473& SiO, v=0-10 & 21 0 -- 20 0 & *\,910.7969& \cr 
& p-H$_2$CO & 14(0,14) -- 13(0,13) & *\,991.6601& o-H$_2$S & 2(2,1) -- 2(1,2) & *\,505.5652& SiO, v=0-10 & 22 0 -- 21 0 &  *\,954.0551& \cr 
& p-H$_2$CO & 14(2,12) -- 13(2,11) & 1\,040.8651& o-H$_2$S & 7(5,2) -- 7(4,3) & *\,555.2540& SiO, v=0-10 & 23 0 -- 22 0 &  *\,997.2976& \cr 
& p-H$_2$CO & 15(2,14) -- 14(2,13) & 1\,085.1676& o-H$_2$S & 5(3,2) -- 5(2,3) & *\,611.4416& SiO, v=0-10 & 24 0 -- 23 0 &  1\,040.5236& \cr 
& o-H$_2$CO & 7(1,7) -- 6(1,6) & *\,491.9684& o-H$_2$S & 4(4,1) -- 4(3,2) & *\,650.3742& SiO, v=0-10 & 25 0 -- 24 0 & 1\,083.7324& \cr 
& o-H$_2$CO & 7(5,2) -- 6(5,1) & *\,509.5621& o-H$_2$S & 3(1,2) -- 3(0,3) & *\,708.4704& SiO, v=0-10 & 26 0 -- 25 0 & 1\,126.9233& \cr 
& o-H$_2$CO & 7(5,3) -- 6(5,2) & *\,509.5621& o-H$_2$S & 2(1,2) -- 1(0,1) & *\,736.0341& SiO, v=0-10 & 27 0 -- 26 0 & 1\,170.0955& \cr 
& o-H$_2$CO & 7(3,5) -- 6(3,4) & *\,510.1558& o-H$_2$S & 4(3,2) -- 4(2,3) & *\,765.9379& SiO, v=0-10 & 28 0 -- 27 0 & 1\,213.2484& \cr 
& o-H$_2$CO & 7(3,4) -- 6(3,3) & *\,510.2378& o-H$_2$S & 3(0,3) -- 2(1,2) & *\,993.1018& SO, v=0 & 7(7) -- 6(7) &     *\,487.7083& \cr 
& o-H$_2$CO & 7(1,6) -- 6(1,5) & *\,525.6658& o-H$_2$S & 5(2,3) -- 5(1,4) & *\,993.1018& SO, v=0 & 4(3) -- 1(2) &     *\,504.6763& \cr 
& o-H$_2$CO & 8(1,8) -- 7(1,7) & *\,561.8993& o-H$_2$S & 6(4,3) -- 6(3,4) & 1\,025.8844& SO, v=0 & 12(11) -- 11(10) & *\,514.8537& \cr 
& o-H$_2$CO & 8(5,4) -- 7(5,3) & *\,582.3821& o-H$_2$S & 4(2,3) -- 4(1,4) & 1\,026.5112& SO, v=0 & 12(12) -- 11(11) & *\,516.3358& \cr 
& o-H$_2$CO & 8(5,3) -- 7(5,2) & *\,582.3821& o-H$_2$S & 2(2,1) -- 1(1,0) & 1\,072.8365& SO, v=0 & 12(13) -- 11(12) & *\,517.3545& \cr 
& o-H$_2$CO & 8(3,6) -- 7(3,5) & *\,583.1446& o-H$_2$S & 3(1,2) -- 2(2,1) & 1\,196.0121& SO, v=0 & 8(8) -- 7(8) &  *\,527.9412& \cr 
& o-H$_2$CO & 8(3,5) -- 7(3,4) & *\,583.3086& CS, v=0-4 & 10 1 -- 9 1 &  *\,486.2010& SO, v=0 & 13(12) -- 12(11) & *\,558.0876& \cr 
& o-H$_2$CO & 8(1,7) -- 7(1,6) & *\,600.3306& CS, v=0-4 & 10 0 -- 9 0 &  *\,489.7509& SO, v=0 & 13(13) -- 12(12) & *\,559.3197& \cr 
& o-H$_2$CO & 9(1,9) -- 8(1,8) & *\,631.7028& CS, v=0-4 & 11 1 -- 10 1 & *\,534.7840& SO, v=0 & 13(14) -- 12(13) & *\,560.1786& \cr 
& o-H$_2$CO & 9(7,3) -- 8(7,2) & *\,654.4633& CS, v=0-4 & 11 0 -- 10 0 & *\,538.6890& SO, v=0 & 9(9) -- 8(9) &     *\,568.7414& \cr 
& o-H$_2$CO & 9(7,2) -- 8(7,1) & *\,654.4634& CS, v=0-4 & 12 0 -- 11 0 & *\,587.6165& SO, v=0 & 14(13) -- 13(12) & *\,601.2584& \cr 
& o-H$_2$CO & 9(5,5) -- 8(5,4) & *\,655.2121& CS, v=0-4 & 13 0 -- 12 0 & *\,636.5324& SO, v=0 & 14(14) -- 13(13) & *\,602.2930& \cr 
& o-H$_2$CO & 9(5,4) -- 8(5,3) & *\,655.2121& CS, v=0-4 & 14 0 -- 13 0 & *\,685.4359& SO, v=0 & 14(15) -- 13(14) & *\,603.0216& \cr 
& o-H$_2$CO & 9(3,7) -- 8(3,6) & *\,656.1647& CS, v=0-4 & 15 0 -- 14 0 & *\,734.3259& SO, v=0 & 10(10) -- 9(10) &  *\,609.9601& \cr 
& o-H$_2$CO & 9(3,6) -- 8(3,5) & *\,656.4646& CS, v=0-4 & 16 0 -- 15 0 & *\,783.2015& SO, v=0 & 5(4) -- 2(3) &     *\,611.5524& \cr 
& o-H$_2$CO & 9(1,8) -- 8(1,7) & *\,674.8098& CS, v=0-4 & 17 0 -- 16 0 & *\,832.0617& & & & \cr
& o-H$_2$CO & 10(1,10) -- 9(1,9) &   *\,701.3704& CS, v=0-4 & 18 0 -- 17 0 & *\,880.9056& & & & \cr
& o-H$_2$CO & 10(5,6) -- 9(5,5) &    *\,728.0535& CS, v=0-4 & 19 0 -- 18 0 & *\,929.7321& & & & \cr
& o-H$_2$CO & 10(5,5) -- 9(5,4) &    *\,728.0536& CS, v=0-4 & 20 0 -- 19 0 & *\,978.5404& & & & \cr
& o-H$_2$CO & 10(3,8) -- 9(3,7) &    *\,729.2126& CS, v=0-4 & 21 0 -- 20 0 & 1\,027.3295& & & & \cr
& o-H$_2$CO & 10(3,7) -- 9(3,6) &    *\,729.7250& CS, v=0-4 & 22 0 -- 21 0 & 1\,076.0984& & & & \cr
& o-H$_2$CO & 10(1,9) -- 9(1,8) &    *\,749.0719& CS, v=0-4 & 23 0 -- 22 0 & 1\,124.8461& & & & \cr
& o-H$_2$CO & 11(1,11) -- 10(1,10) & *\,770.8961& CS, v=0-4 & 24 0 -- 23 0 & 1\,173.5718& & & & \cr
& o-H$_2$CO & 11(5,7) -- 10(5,6) &   *\,800.9075& CS, v=0-4 & 25 0 -- 24 0 & 1\,222.2744& & & & \cr
& o-H$_2$CO & 11(3,9) -- 10(3,8) &   *\,802.2824& CS, v=0-4 & 26 0 -- 25 0 & 1\,270.9529& & & & \cr
& o-H$_2$CO & 11(3,8) -- 10(3,7) &   *\,803.1116& $^{13}$CS, v=0,1 & 11 0 -- 10 0 & *\,508.5347& & & & \cr
& o-H$_2$CO & 11(1,10) -- 10(1,9) &  *\,823.0828& $^{13}$CS, v=0,1 & 12 0 -- 11 0 & *\,554.7257& & & & \cr
& o-H$_2$CO & 12(1,12) -- 11(1,11) & *\,840.2757& $^{13}$CS, v=0,1 & 13 0 -- 12 0 & *\,600.9065& & & & \cr
& o-H$_2$CO & 12(3,10) -- 11(3,9) &  *\,875.3662& $^{13}$CS, v=0,1 & 15 0 -- 14 0 & *\,693.2337& & & & \cr
& o-H$_2$CO & 12(3,9) -- 11(3,8) &   *\,876.6491& $^{13}$CS, v=0,1 & 16 0 -- 15 0 & *\,739.3785& & & & \cr
& o-H$_2$CO & 12(1,11) -- 11(1,10) & *\,896.8051& $^{13}$CS, v=0,1 & 17 0 -- 16 0 & *\,785.5096& & & & \cr
& o-H$_2$CO & 13(1,13) -- 12(1,12) & *\,909.5077& $^{13}$CS, v=0,1 & 18 0 -- 17 0 & *\,831.6261& & & & \cr
&\multispan9\hrulefill& \cr
\noalign{\vspace{-7.5pt}}
}
\end{center}
\endgroup
\vspace{-12pt}
\end{table*}


\bsp	
\label{lastpage}
\end{document}